\documentclass[a4paper,11pt]{article}
\pdfoutput=1 

\usepackage{jheppub} 

\usepackage{bm,amsmath,amssymb,slashed,graphicx,%
            enumerate,alltt,xspace,multirow,xcolor,mathrsfs}
\usepackage{fancyvrb}
\usepackage{cancel}

\usepackage[normalem]{ulem}

\usepackage{subcaption}

\RequirePackage{braket}
\RequirePackage{graphicx}

\RequirePackage[numbers,sort&compress]{natbib}

\makeatletter
\g@addto@macro\bfseries{\boldmath}
\makeatother

\definecolor{labelkey}{rgb}{0,0.5,0.0}

\usepackage{listings}
\definecolor{darkgreen}{rgb}{0,0.4,0}
\newcommand\nf{n_f} 

\newcommand\zetatj{\zeta_{\scriptscriptstyle 2J}}
\newcommand\Mz{M_{\scriptscriptstyle \rm Z}}
\newcommand\Mh{M_{\scriptscriptstyle \rm H}}
\newcommand\Md{M_{\scriptscriptstyle \rm D}}
\newcommand\mur{\mu_{\rm r}}

\newcommand\mathd{\mathrm{d}}

\newcommand{\as}{\alpha_s}

\newcommand{\Bw}{B_W}

\newcommand{\muR}{\mu_{\scriptscriptstyle \rm R}}
\newcommand{\xR}{x_{\scriptscriptstyle \rm R}}

\newcommand{\Pythiaeight}{{\tt Pythia8{}}}
\newcommand{\PythiaEight}{{\tt Pythia8 {}}}
\newcommand{\HerwigSix}{{\tt Herwig6 {}}}
\newcommand{\HerwigSeven}{{\tt Herwig7 {}}}

\newcommand{\Hnp}{H_{\rm \scriptscriptstyle NP}}
\newcommand{\Ca}{C_{\rm \scriptscriptstyle A}}
\newcommand{\Cf}{C_{\rm \scriptscriptstyle F}}

\newcommand{\SigmaLO}{\Sigma_{B}}
\newcommand{\SigmaFULL}{\Sigma_{\rm \scriptscriptstyle FULL}}

\title{\mbox{Fits of $\alpha_s$ using power corrections in the three-jet region}}
\preprint{
  \begin{flushright}
    MPP-2022-142
  \end{flushright}
}

\author[a,b]{Paolo Nason,}
\author[b,c]{Giulia Zanderighi}

\emailAdd{paolo.nason@mib.infn.it}
\emailAdd{zanderi@mpp.mpg.de}

\affiliation[a]{Universit\`a di Milano-Bicocca and INFN, Sezione di
  Milano-Bicocca, Piazza della Scienza 3,20126 Milano, Italy}
\affiliation[b]{Max-Planck-Institut f\"ur Physik, F\"ohringer Ring 6,
  80805 M\"unchen, Germany} \affiliation[c]{Physik-Department,
  Technische Universit\"at M\"unchen, James-Franck-Strasse 1, 85748
  Garching, Germany}

\date{Received: date / Accepted: \today}

\abstract{In this work we study the impact of recent findings
  regarding non-perturbative corrections in the three-jet region to
  $e^+e^-$ hadronic observables, by
  performing a simultaneous fit of the strong coupling constant
  $\alpha_s$ and the non-perturbative parameter $\alpha_0$.
  We extend the calculation of these power corrections, already known
  for thrust and C-parameter, to other $e^+e^-$ hadronic
  observables. We find that for some observables the non-perturbative
  corrections are reasonably well behaved in the two-jet limit, while
  for others they have a more problematic behaviour.
  If one limits the fit to the three-jet region and to the
  well-behaved observables, one finds in general very good results,
  with the extracted value of $\alpha_s$ agreeing well with
  the world average. This is the case in particular for the thrust and
  $C$-parameter for which notably small values of $\alpha_s$ have been
  reported when non-perturbative corrections have been computed using
  analytic methods. Furthermore, the more problematic variables are
  also well described provided one stays far enough from the two-jet
  limit, while in this same region they cannot be described using the
  traditional implementation of power-corrections based on two-jet
  kinematics.}


\begin{document}

\maketitle


\newcommand{\citere}[1]{Ref.\,\cite{#1}}
\newcommand{\citeres}[1]{Refs.\,\cite{#1}}

\section{Introduction}
\label{sec:intro}
The study of shape variables in $e^+e^-$ annihilation is one of the simplest
contexts in which to test perturbative QCD, and it is potentially among the
cleanest frameworks where one can measure the strong coupling constant
 $\alpha_s$ at high energy by probing directly the
quark-antiquark-gluon vertex. Shape variables have been computed up to
order
$\as^3$~\cite{Gehrmann-DeRidder:2007foh,Gehrmann-DeRidder:2007vsv,Gehrmann-DeRidder:2008qsl,DelDuca:2016csb},
and resummations near the two-jet region have been performed at
different levels of accuracy, either using traditional resummation
methods~\cite{Catani:1992ua,Catani:1998sf,Dokshitzer:1998kz,Banfi:2001bz,Monni:2011gb,Banfi:2014sua,Tulipant:2017ybb,Banfi:2016zlc},
or using Soft Collinear Effective Theory
(SCET)~\cite{Becher:2008cf,Chien:2010kc,Becher:2012qc,Becher:2011pf},
leading to very precise predictions at high energies.

It is well known, however, that shape variables are affected by
linearly suppressed power corrections, i.e.\ of the order of
$\Lambda/Q$, where $\Lambda$ is a typical hadronic scale and $Q$ is
the annihilation energy. Since in the 3-jet region the shape variables
are of order $\as$, this implies a relative error of order
$(\Lambda/Q)/\as$, that affects at the same level the measured value
of $\as$.  If we assume that $\Lambda$ is of the order of $0.5$~GeV
(i.e.\ the typical additional transverse energy per unit of rapidity
due to hadronization), on the $Z$ peak we estimate an error of the order of 5\%. In
practice, power corrections can reach the 10\% level for some
observables.

A commonly adopted approach for dealing with power corrections in the
determinations of the strong coupling constant from shape variables is
to use Monte Carlo
models~\cite{Dissertori:2009ik,OPAL:2011aa,Bethke:2008hf,Dissertori:2009qa,Schieck:2012mp,Verbytskyi:2019zhh,Kardos:2018kqj}.
A shower Monte Carlo is used to construct a migration matrix for shape
variables computed from final-state hadrons, and from partons before
hadronization. The migration matrix is then applied to the measured
differential distribution of hadrons to obtain the shape distribution
in terms of partons. This is in turn compared to perturbative QCD, and
a value of $\as$ is extracted.  This method is often criticized,
because the Monte Carlo hadronization model does not bear a clean
relation to field-theoretical calculations.

An alternative strategy for the inclusion of power corrections makes
use of analytic approaches.  In this case, the theoretical calculation
including power corrections is compared directly to the shape variable
measurement using hadrons. These methods can be classified into two
broad classes.

One approach makes use of an effective coupling for the emission of
very soft gluons (called ``gluers'')
\cite{Akhoury:1995sp,Dokshitzer:1995qm,Dokshitzer:1997iz,Dokshitzer:1998pt}.
The average value of the effective coupling in a given low-energy
range plays the role of a parameter to be fitted to data together with
the value of $\as$.  The Particle Data
Group~\cite{ParticleDataGroup:2020ssz} (PDG) currently includes two
fits of the strong coupling based on NNLO+NLL~\cite{Davison:2008vx} or
NNLO+NNLL~\cite{Gehrmann:2012sc} accurate perturbative results
combined with this approach to the non-perturbative corrections.

This approach is also motivated by the large-$\nf$ limit of QCD~(see
\cite{Beneke:1998ui} and references therein), where the effective
coupling can be actually computed. It is argued that the
non-perturbative parameter in this contest is universal, i.e.\ it is
the same for a large class of shape variables.  The coefficient of the
power correction is computed by simply adding a gluer to an initial
$q\bar{q}$ state.  For shape variables that are additive in soft
radiation near the two jet limit, the emission of the gluer acts as a
shift in the value of the shape variable. This behaviour is then
extrapolated to the three-jet region, i.e. the non perturbative
correction is included as a shift in the argument of the shape
variable computed in perturbation theory.

The other approach relies upon factorization in
QCD~\cite{Korchemsky:1999kt,Korchemsky:2000kp,Bauer:2003di,Lee:2006nr}.
This begins with the computation of the shape variables including
resummation of the soft-collinear singularities arising from gluon
emission from the primary quark and antiquark. The region of very soft
emissions is parameterized by a shape function that is factorized out
of the distribution.  In the three-jet region, a single moment of the
shape function controls the linear non-perturbative corrections. This
approach arises naturally in SCET~\cite{Bauer:2000yr,Bauer:2001yt}.
Two determinations of $\as(M_Z)$ included in the
PDG~\cite{Abbate:2010xh,Hoang:2015hka} currently rely on such analytic
SCET-based approaches and notably lead to low values of the strong
coupling accompanied by small uncertainties.

A common feature of these two approaches is that they rely upon the
extrapolation of the non-perturbative correction from the two-jet to
the three-jet limit. This extrapolation has been shown not to agree
with the direct calculation of the non-perturbative correction for the
$C$ parameter near the three-jet symmetric
limit~\cite{Luisoni:2020efy}, where it leads to an overestimate by
approximately a factor of two.

In refs.~\cite{Caola:2021kzt,Caola:2022vea} it was shown that linear
power corrections in the bulk of the three parton final state region
can be computed in large-$\nf$ QCD in the process $e^+e^-\to q\bar{q}
\gamma$, and, under some further assumptions, also in the $e^+e^-\to
q\bar{q} g$ process. In ref.~\cite{Caola:2022vea} the result for the
$C$-parameter and thrust was given, but the method is quite general
and can be extended to a wide class of shape variables. In the case of
the $C$-parameter it leads to a result consistent with
ref.~\cite{Luisoni:2020efy} in the three-jet symmetric limit. In
general as for the $C$-parameter case, one finds considerable
violations of the assumption that the non-perturbative correction can
be implemented as a constant shift of the perturbative result well
into the three-jet region.

The purpose of this work is to investigate whether there are some
indications that the newly computed power corrections are preferred by
available data. In order to do this, we considered $Z$-peak data from
the ALEPH experiment~\cite{ALEPH:2003obs} that are publicly available
on HEPDATA and quite precise, and consider a set of shape variables
such that the computation along the lines of ref.~\cite{Caola:2022vea}
can be carried out.
Besides thrust and the C-parameter, ref.~\cite{ALEPH:2003obs} provides
data for other shape variables for which we are in a position to
compute non-perturbative corrections in the three-jet region, namely
the square mass of the heavy hemisphere $\Mh^2$, the difference of the
squares masses of the heavy and light hemisphere $\Md^2$, the
broadening of the wide jet $B_W$, and the 3-jet resolution parameter
$y_3$ in the Durham scheme.
In this work we have then computed the non-perturbative coefficients for
$\Mh^2$, $\Md^2$, $B_W$, and, with some caveats to be detailed in the
following, also for $y_3$. We thus supplement the $\as^3$ calculation
of these shape variables with the inclusion of the non-perturbative
corrections that we have computed as a shift in the argument of the
cumulative cross section~$\Sigma(v)$.  More precisely, calling $V$ a generic shape
variable, defined in such a way that it vanishes in the two jet limit,
$\Sigma(v)$ is defined as the cross section for producing
events such that $V<v$.  In our approach, the shift in the argument is
given by $v\to v-\zeta(v) \Hnp$, where $\Hnp$ is a coefficient
suppressed by a power of $Q$, equal for all shape variables, and
$\zeta(v)$ is a shape-variable specific, dimensionless function.  In
contrast, in the traditional form of the power corrections the
variable-specific function $\zeta(v)$ is evaluated in the two-jet
limit, where it is in most cases replaced by a constant.\footnote{In
  the case of the broadening the shift is not a constant, see
  ref.~\cite{Dokshitzer:1998qp}.}

We stress that, somewhat unconventionally, we do not include
resummation effects in our result, while it is common practice to
include them also very far away from the two-jet limit.  They
generally lead to an increase of the shape variable distributions,
and thus to a smaller value of $\as$. We take here the point of view
that if we consider ranges of the shape variables that are far enough
from the 2-jet limit, resummation can be neglected. The reader may
keep in mind that if resummation effects were included we would
generally obtain smaller values of $\as$.

A further reason for not including resummation in our result is that
it is not clear whether including the constant non-perturbative shifts
in the singular contributions is an acceptable procedure. In fact,
such corrections would propagate into the three-jet region, where (as we
will see later)
they sharply differ from their two-jet limit.
Furthermore, in this work we will not try to give a preferred value of
$\as$ with an error. Rather, our aim is only to see whether and where
the newly computed non-perturbative corrections are in some way
preferred by data, and to assess their impact.

The rest of the paper is organized as follows. In
Sec.~\ref{sec:shapevars} we define the observables that we consider in
this work. In Sec.~\ref{sec:masses} we discuss ambiguities in the
event-shape definitions that arise when dealing with massive hadrons,
as opposed to massless QCD partons, and recall three alternative
definitions that differ for massive hadrons but
agree for massless partons. In Sec.~\ref{sec:powcor} we present the
calculation of the power-corrections in the three-jet region for all
observables considered in this work and show that they give rise to
a non-constant shift of the perturbative distribution. We also discuss
numerical checks of the analytic calculations. In
Sec.~\ref{sec:obscalc} we discuss how to combine perturbative ${\cal
  O}(\alpha_s^3)$ results with non-perturbative corrections. In
particular, we define various schemes that differ by higher order
terms. In Sec.~\ref{sec:fitALEPH} we discuss our treatment of
uncertainties and correlations, as well as the corrections
that we apply to account for the heavy-quark masses.
Finally, in Sec.~\ref{sec:fit} we present the results of our fits of
$\alpha_s$. We discuss various ambiguities and uncertainties, as well
as their difference from fits relying on the calculation of
non-perturbative corrections in the two-jet region. We conclude in
Sec.~\ref{sec:conclu}. In App.~\ref{app:resum} we discuss the impact
of all-order resummation effects for the observables used in our fit.

\section{Observable definitions}\label{sec:shapevars}

The choice of event shapes considered in this work is based on whether
ALEPH data are available for them, and whether their associated
non-perturbative corrections in the three jet region can be calculated
along the lines of ref.~\cite{Caola:2022vea}, as discussed in detail
in Sec.~\ref{sec:powcor}.
Unless otherwise specified, all sums in the definitions below run over
all particles in the event.

\begin{itemize}
\item {\bf The thrust $T$}, or $\tau = 1-T$, is defined as
\begin{equation}
T = \max_{{\vec n}_T}\left( \frac{\sum_i |\vec p_i\cdot \vec n_T|}{\sum_i |\vec p_i|}\right)\,,
\end{equation}
where the axis $\vec n_T$, that maximises the sum, is the thrust axis
of the event. 
\item {\bf The heavy-jet mass:} the plane through the origin of the
  event, orthogonal to the thrust axis $\vec n_T$, divides each event
  into two hemispheres ${\cal H}_j$ ($j=1,2$), the invariant mass of
  each is defined as
  \begin{equation}
    M_j^2 = \frac{1}{E_{\rm vis}^2}\left(\sum_{p_i \in {\cal H}_j} p_i
    \right)^2\,,\quad j=1,2\,,
  \end{equation}
  where $E_{\rm vis} = \sum_i E_i$. 
  The heavy-jet mass is the larger of the two
\begin{equation}
    \Mh^2 = \max\left(M_1^2,M_2^2\right)\,. 
  \end{equation}
\item {\bf The jet mass difference} is defined as the difference
  between the larger and smaller of the two masses
\begin{equation}
    \Md^2 = |M_1^2-M_2^2|\,. 
\end{equation}

\item {\bf The $C$-parameter} is computed from the three eigenvalues
  $\lambda_i$ of the momentum tensor $\Theta^{\alpha\beta}$
  \begin{equation}
\Theta^{\alpha\beta} = \frac{1}{\sum_i |p_i|} \sum_i \frac{p_i^\alpha p_i^\beta}{|\vec p_i|}, \quad \alpha, \beta = 1,2,3\,,
  \end{equation}
  as
  \begin{equation}
C = 3 \cdot \left( \lambda_1 \lambda_2+ \lambda_1 \lambda_3+ \lambda_2
\lambda_3\right)\,.
\label{eq:cpar}
  \end{equation}  

\item {\bf The wide broadening:} given the thrust axis $n_T$, the
  hemisphere broadenings $B_j$ ($j=1,2$) measure the amount of
  transverse momentum in each hemisphere
    \begin{equation}
    B_j = \frac{\sum_{p_i \in H_j} | \vec p_i \times \vec n_T|}{2 \sum_i |\vec p_i|}\,,\quad j=1,2\,.    
\end{equation}
    The wide broadening $B_W$ is the larger of the two hemisphere broadenings
    \begin{equation}
    B_W = \max\left(B_1,B_2\right)\,. 
\end{equation}    
\item {\bf The three-jet resolution $y_3$:} we take the Durham jet
  clustering, whose distance measure reads
  \begin{equation}
    y_{ij} = \frac{2 \min(E_i^2,E_j^2)(1-\cos\theta_{ij})}{E_{\rm vis}^2}\,.
  \end{equation}
(Pseudo)-jets are recombined sequencially summing the four-momenta of
  the pair of particles with the smallest $y_{ij}$.  The three-jet
  resolution $y_3$ is defined as the value of $y_{\rm cut}$ for which
  an event changes from being classified as $2$- to $3$-jet.
\end{itemize}
The published data are already corrected using Monte Carlo generators
in such a way that all particles produced by the $e^+e^-$ annihilation
are included, comprising also the neutrinos from meson decays.

\section{Hadron mass ambiguities}\label{sec:masses}
When computing shape variables in perturbative QCD, one always deals
with massless partons. However, the measurements use the
four-momenta of massive hadrons. It turns out that shape variable
definitions may differ for massive hadrons and be identical for
massless partons, and this introduces an ambiguity in the experimental
definition of the event shapes. This problem has been studied in
detail in ref.~\cite{Salam:2001bd} (see also \cite{Mateu:2012nk}),
where three alternative schemes where suggested: the $p$-scheme, the
$E$-scheme and the $D$-scheme.
In the $p$-scheme one uses only the three-momenta of the particles
$\vec{p}_i$, and the energies $E_i$ are replaced by $|\vec p_i|$.
Instead, in the $E$-scheme the energies of the particles are
preserved, but the three-momenta are rescaled so as to have massless
four-momenta, $\vec p_i \to \vec p_i\cdot \frac{E_i}{|\vec p_i|}$.

It is clear that in the $p$-scheme energy conservation is violated,
while in the $E$-scheme the three momentum is not conserved, the
violation being in both cases of the order of the hadron masses.

In the so called $D$-scheme, final state hadrons are decayed
isotropically in their rest frame into two fictitious massless
particles. The event shape is then computed using only massless
particles. This scheme has the advantage that the full four-momentum
of the event is conserved, and that no reference to a particular frame
needs to be invoked in its implementation.  Notice also that it can
happen that long-lived, unstable hadrons are produced that decay to
lighter particles. Therefore the event shape depends on the level at
which the measurement is performed, i.e. it becomes relevant whether
the measurement is performed before or after these decays. Unlike all
other schemes, the $D$-scheme has the advantage that it is rather
insensitive to the particular hadron level chosen to perform the
measurement~\cite{Salam:2001bd}.

In ref.~\cite{Salam:2001bd}, the advantages and disadvantages of each
of these schemes are discussed. In particular it is argued that in the
$E$-scheme non-universal mass effects are absent. The arguments used
there are based upon an analysis near the two-jet limit, and their
applicability to the case of three widely separated jets is unclear.
One may also argue that the $D$-scheme should be preferred, since it
mimics to some extent the models of hadron formation.  In the present
work we will adopt the $E$-scheme as our default choice and use the
additional three schemes to gauge the hadron-mass sensitivity of our
results.

\section{Power correction calculation}
\label{sec:powcor}
According to ref.~\cite{Caola:2022vea}, provided an event shape
satisfies specific conditions, as explained in detail later, its power
correction in the three-jet region can be computed according to
the formula
\begin{equation}\label{eq:npcorr}
  \left[\Sigma(v)\right]_{\rm NP}=\left\{\int \mathd \sigma_B(\Phi_B) \delta(v(\Phi_B) -v)
  \sum_{\rm dip}\left[-{\cal M} \times 4 \frac{\as C_{\rm dip}}{2\pi}\frac{1}{Q}
    \int \mathd \eta\,\frac{\mathd \phi}{2\pi} h_v(\eta, \phi)\right]\right\}
   \times I_{\rm NP},
 \end{equation}
 where $Q$ is the total center of mass (CM) energy and the sum runs over
 all radiating dipoles associated with the given Born configuration.
 Thus, for the two jet case there is just a single $q\bar{q}$ dipole,
 while for the three-jet case we have a $q\bar{q}$, $q g$ and $\bar{q}
 g$ dipole.\footnote{The same formula is also applicable to higher
   multiplicity Born processes, that we do not consider here.} We
 stress that the function $h_v$ depends also upon $\Phi_B$, and that
 for ease of notation we do not show explicitly this dependence. The
 colour coefficients $C_{\rm dip}$ for the three-jet case are given by
 \begin{equation}
   C_{q\bar{q}}= \Cf-\frac{\Ca}{2},\quad\quad C_{qg}=C_{\bar{q}g}=\frac{\Ca}{2}.
 \end{equation}
 The Milan factor ${\cal M}$ is given in analytic form in ref.~\cite{Smye:2001gq}
 \begin{equation}\label{eq:mifac}
   {\cal M}= \frac{3}{64} \frac{(128\pi(1+\log 2)-35\pi^2) \Ca
     -10\pi^2 T_R n_F}{11 \Ca-4T_R n_F},
 \end{equation}
 that agrees with the numerical result given earlier in ref.~\cite{Dokshitzer:1998pt}
 \begin{equation}\label{eq:mifacnumeric}
   {\cal M} =
   1+(1.575\Ca - 0.104 n_f)/\beta_0,
 \end{equation}
 where $\beta_0 = (11\Ca - 4 n_f T_R)/3$.
 The coefficient $I_{\rm NP}$ depends upon the model used to implement
 power corrections. In the large-$n_F$ theory, it has the expression
 (see e.g. Ref.~\cite{FerrarioRavasio:2018ubr})
 \begin{align}\label{eq:largenfINP}
   I_{\rm NP}&=\frac{1}{b_{0,\nf}\as(\mu)}\int_0^{\mu_C}\frac{d\lambda}{\pi} \arctan \frac{\pi b_{0,\nf} \as(\mu)}{1+b_{0,\nf} \as(\mu)\log \frac{\lambda^2e^{-5/3}}{\mu^2}} \nonumber \\
   &=\frac{1}{\as(\mu)}\int_0^{\mu_C}\mathd\lambda \frac{\arctan({\pi b_{0,\nf} \as(\lambda e^{-5/6})})}{\pi b_{0,\nf}},
 \end{align}
 where $b_{0,\nf}=-\nf/(6\pi)$.  The upper limit of integration in
 eq.~\eqref{eq:largenfINP} is quite arbitrary. It should be large
 enough to cover the region where the argument of
 the arctangent diverges, corresponding to the Landau pole.  In
 phenomenological models $I_{\rm NP}$ is replaced by the integral over
 a non-perturbative effective coupling, given as function of a scale
 $\lambda$.

 The function $h_v(\eta,\phi)$ depends upon the shape variable. It is
 defined as
 \begin{equation}
   h_v(\eta,\phi)=\lim_{|l_\perp| \to 0}\frac{1}{|l_\perp|}
   \left(v(\left\{P\right\},l)-v(\left\{p\right\})\right),
 \end{equation}
 where $\left\{P\right\}$ denote the momenta of the hard final state
 partons after the radiation of a soft massless parton of momentum $l$,
 and $\left\{p\right\}$ denote the momenta of the final state partons in
 the absence of radiation. The arguments $\eta$ and $\phi$ are the
 rapidity and azimuth of the soft parton, and $l_\perp$ denotes its
 transverse momentum, all evaluated in the rest
 frame of the radiating dipole.
 The mapping from $\left\{P\right\}$ and $l$
 to $\left\{p\right\}$ must have certain smoothness properties, namely
 the momenta $\left\{P\right\}$ must be functions of
 $\left\{p\right\}$ and $l$ that are linear in $l$ for small $l$.
 
 There are two further requirements for formula~(\ref{eq:npcorr}) to hold.
 The first one is that it applies to variables that are additive in
 the emission of more than one soft parton in the three-jet region.
 This property is violated
 by $y_3$, as discussed in the following.  The second one is that the
 function $h_v(\eta,\phi)$, after azimuthal integration, should yield
 a convergent integral in $\eta$. This property is violated, for
 example, by the total broadening, and that is the reason why we do
 not consider it in this work.
 
 Notice that in the large $\nf$ limit the Milan factor becomes equal
 to $15\pi^2/128$, and the expression in the curly bracket of
 eq.~(\ref{eq:npcorr}) becomes equal to eq.~(4.7) of
 Ref.~\cite{Caola:2022vea}, up to the $\lambda$ factor. In fact,
 according to eq.~(A.1) of ref.~\cite{Caola:2022vea}, the linear
 non-perturbative correction to an observable in the large $\nf$ limit
 is proportional to the first order coefficient of its expansion in
 $\lambda$, where $\lambda$ is a (fictitious) gluon mass introduced in
 the calculation, multiplied by the factor given in
 eq.~(\ref{eq:largenfINP}).

 In ref.~\cite{Caola:2022vea} the integration in $\eta$ and $\phi$ was
 performed analytically for the $C$-parameter and for thrust. Here we
 have set up a numerical code to perform the $\eta$ and $\phi$
 integration numerically, since a sufficient precision can be easily
 reached, and this allows us to add more observables with relatively
 minor effort.
 
\subsection{Thrust}
 We illustrate now how the
 $h_v(\eta,\phi)$ function is computed in our code, using thrust
 as an example.
 We generate the Born momenta $\left\{p\right\}$ according to the
 three-body phase space.  Let us assume for definiteness that $p_1$,
 $p_2$ is the radiating dipole.  We generate $\eta$ and $\phi$, and
 construct the four-vector
 \begin{align}
   l&=l^++l^-+l_\perp, \nonumber \\ l^+& =\frac{p_1}{\sqrt{2p_1\cdot
       p_2}} \exp(\eta), \nonumber \\ l^-&= \frac{p_2}{\sqrt{2p_1\cdot
       p_2}}\exp(-\eta), \label{eq:ldef}
 \end{align}
 where
 \begin{equation} \label{eq:ldefprops}
   l_\perp\cdot l^+=0,\quad\quad l_\perp\cdot l^-=0, \quad\quad l_\perp^2=-1,
 \end{equation}
 and $l_\perp$ has an azimuthal angle $\phi$ relative to the $p_{1/2}$
 axis in the dipole rest frame.  Notice that by construction $l^2=0$.
 
 Let us call $\vec{t}_0$ the trust axis in the CM frame, defined to
 have the direction of the largest $\vec{p}_i$, $i=1\ldots 3$. The
 thrust variation due to the emission of a parton with momentum
 $\lambda l$ is given by
 \begin{equation}
 \delta\tau=-   \delta T=-\frac{1}{Q} \left[ \max_{\vec{t}}\left( \sum_{i=1,3}| \vec{P}_i\cdot \vec{t}|
       + \lambda|\vec{l}\cdot \vec{t}|\right) -\sum_{i=1,3}| \vec{p}_i\cdot \vec{t}_0|\right].
 \end{equation}
 We need to expand this expression for small $\lambda$, keeping only the linear terms. We have three terms
 \begin{align} \nonumber 
   \delta T&= \frac {\sum_{i=1,3}| (\vec{p}_i+\delta \vec{P}_i) \cdot \vec{t}_0| -\sum_{i=1,3}| \vec{p}_i\cdot \vec{t}_0|}{Q} \\    \nonumber 
           &+ \frac {\sum_{i=1,3}| \vec{p}_i \cdot (\vec{t}_0+\delta\vec{t})| -\sum_{i=1,3}| \vec{p}_i\cdot \vec{t}_0|}{Q} \\  
           &+ \lambda \frac { |\vec{l}\cdot \vec{t}_0|}{Q}.   \label{eq:recoil}
 \end{align}
 The second line of eq.~(\ref{eq:recoil}) can be worked out as
 follows. We must have $\delta \vec{t} \cdot \vec{t}_0 = 0$, since
 $\vec{t}$ has fixed length. Thus we have $\delta \vec{t} \cdot
 \vec{p}_k=0$, where $k$ is the hardest parton. For the remaining two
 partons, with $i,j\neq k$ we have
 \begin{equation}
   |\vec{p}_i \cdot (\vec{t}_0+\delta\vec{t})|=|\vec{p}_i \cdot
   \vec{t}_0| \times\left(1+\frac{\delta\vec{t}\cdot \vec
     p_i}{\vec{p}_i \cdot \vec{t}_0}\right) = |\vec{p}_i \cdot
   \vec{t}_0| - \delta\vec{t}\cdot \vec p_i,
 \end{equation}
 where we have used the fact that $\vec{p}_i \cdot \vec{t}_0<0$ for
 $i\neq k$. Thus
 \begin{equation}
   \frac {\sum_{i=1,3}| \vec{p}_i \cdot (\vec{t}_0+\delta\vec{t})| -\sum_{i=1,3}| \vec{p}_i\cdot \vec{t}_0|}{Q}
   = - \frac{1}{Q}\delta\vec{t}\cdot (\sum_{i\neq k} \vec{p}_i) =  \frac{1}{Q}\delta\vec{t}\cdot \vec{p}_k = 0,
 \end{equation}
 since $\delta \vec{t}$ is orthogonal to $\vec{t}_0$ and thus to
 $\vec{p}_k$.  Thus only the terms in the first and last line of
 eq.~(\ref{eq:recoil}) contribute.
 The first line is linear in
 $\delta{P}_i$, and thus (in an appropriate recoil scheme) also in
 $l$. It must have the form
 \begin{equation}
   \frac{\lambda}{Q}(A \exp(\eta) + B \exp(-\eta) + C \sin\phi + D\cos\phi)
 \end{equation}
 with $A$, $B$, $C$ and $D$ depending only upon the Born kinematics.
 The full result is
 \begin{equation}
   \delta T = \frac{\lambda}{Q}( |\vec{l}\cdot \vec{t}_0|+A \exp(\eta) + B \exp(-\eta) + C \sin\phi + D\cos\phi).
 \end{equation}
 The above expression must however not lead to a divergent integral
 for large rapidity. Looking, for example, at the large $\eta$ limit
 of the above expression (see eqs.~(\ref{eq:ldef})), we have
 \begin{align}
   |\vec{l}\cdot \vec{t}_0|&=\left|\frac{\vec{p}_1\cdot \vec{t}_0}{\sqrt{2 p_1\cdot p_2}} \exp(\eta)+\frac{\vec{p}_2\cdot \vec{t}_0}{\sqrt{2 p_1\cdot p_2}}\exp(-\eta)
                             +\vec{l}_\perp \cdot \vec{t}_0 \right| \\
                           &=\left|\frac{\vec{p}_1\cdot \vec{t}_0}{\sqrt{2 p_1\cdot p_2}}\right|
                             \left(\exp(\eta)+\frac{\vec{p}_2\cdot \vec{t}_0}{\vec{p}_1\cdot \vec{t}_0 }\exp(-\eta)
                             + \frac{\vec{l}_\perp \cdot \vec{t}_0}{\vec{p}_1\cdot \vec{t}_0} \right)
 \end{align}
 We thus see that by choosing
 \begin{equation}
   A = - \left|\frac{\vec{p}_1\cdot \vec{t}_0}{\sqrt{2 p_1\cdot p_2}}\right|
 \end{equation}
 we cancel that exponential growth in $\eta$. With an analogous choice
 for $B$ we can cancel the exponential divergence for $\eta \to
 -\infty $. Terms with constant behaviour for large $\eta$ do remain,
 but they cancel after azimuthal integration.  Thus, our final
 expression for the $h_v$ function for $\tau$ is obtained by changing
 sign to the previous expression,
 \begin{equation}
   h_\tau(\eta,\phi) = -h_T(\eta,\phi)= -|\vec{l}\cdot \vec{t}_0| + |\vec{l}^+\cdot \vec{t}_0| + |\vec{l}^-\cdot \vec{t}_0|.
 \end{equation}
 In order to explicitly get rid of the constant $\phi$-dependent term,
 in the numerical integration process we sum the two contributions
 obtained with the replacement $\phi\to \phi+\pi$.

 \subsection{Other observables}\label{sec:other}
With a similar procedures we find the expression of $h_v$ for all shape
variables of our interest, for which we report here only the final
results.  For the $C$ parameter, starting from the expression
\begin{equation}
C=3-\frac{3}{2}\sum_{i,j}\frac{(p_i\cdot p_j)^2}{(p_i\cdot q)(p_j\cdot q)},
\end{equation}
valid for massless partons, we obtain
\begin{equation}\label{eq:cparh}
  h_C(\eta,\phi) = - 3 \sum_{i=1}^3 \left[ \frac{(l\cdot p_i)^2}{l\cdot q\,p_i\cdot q}
    -\frac{(l^+\cdot p_i)^2}{l^+\cdot q\,p_i\cdot q}
    -\frac{(l^-\cdot p_i)^2}{l^-\cdot q\,p_i\cdot q} \right],
\end{equation}
where $q = \sum_i p_i$. 
The negative terms in the square bracket of eq.~(\ref{eq:cparh}) are
there to cancel the divergent rapidity behaviour of the positive term,
and are clearly linear in the momentum components of $l$.

For the heavy-jet mass we find 
\begin{align}
  h_{\Mh^2}(\eta,\phi)=&
  \theta(t_0\cdot l) \left[ \frac{q\cdot l}{Q}(2-T_0) - T_0\,t_0\cdot l \right]
    -\theta(t_0\cdot l^+) \left[  \frac{q\cdot l^+}{Q}(2-T_0) - T_0\, t_0 \cdot l^+ \right] \nonumber \\
    &-\theta(t_0\cdot l^-) \left[  \frac{q\cdot l^-}{Q}(2-T_0) - T_0\, t_0 \cdot l^- \right]\,,
\end{align}
where $T_0$ stands for the value of the thrust at Born level, and, as
before, the vector $t_0$ is obtained by adding a zero time-component
to the thrust three-vector.  Also in this case the subtraction terms
are clearly identified.  Notice that the theta functions involving
$l^+$ and $l^-$ are actually independent upon $l$, since
\begin{equation}
  \theta(\pm t_0 \cdot l^{+/-})= \theta(\pm t_0 \cdot p_{1/2})\,.
\end{equation}

The light jet mass is given by
\begin{align}
  h_{M_l^2}(\eta,\phi)=&
    \theta(-t_0\cdot l) \, T_0\left[t_0\cdot l+  \frac{q\cdot l}{Q}\right] \nonumber \\
    &- \theta(-t_0\cdot l^+)\,T_0 \left[ t_0\cdot l^++  \frac{q\cdot l^+}{Q} \right]
    - \theta(t_0\cdot l^-)\, T_0\left[ t_0\cdot l^-+  \frac{q\cdot l^-}{Q}\right].
\end{align}
From the heavy- and light-jet mass we also obtain the mass difference
\begin{equation}
 h_{\Md^2}(\eta,\phi)=h_{\Mh^2}(\eta,\phi)-h_{M_l^2}(\eta,\phi).
\end{equation}

The wide jet broadening is given by
\begin{align}
  h_{\Bw}(\eta,\phi)
  =&
     \theta(t_0\cdot l)\frac{1}{2} \sqrt{\left(\frac{l\cdot q}{Q}\right)^2-(t_0\cdot l)^2}
     -\theta(t_0\cdot l^+)\frac{1}{2} \sqrt{\left(\frac{l^+\cdot q}{Q}\right)^2-(t_0\cdot l^+)^2}
     \nonumber \\
   & -\theta(t_0\cdot l^-)\frac{1}{2} \sqrt{\left(\frac{l^-\cdot q}{Q}\right)^2
     -(t \cdot l^-)^2}
      \nonumber \\ \label{eq:bw}
   & + \sum_{i=1}^3 \theta(t\cdot p_i)\theta(-t \cdot  l)
     \frac{
     ( \vec{l}\cdot\vec{p}_i) t\cdot p_i
     - l \cdot t (t \cdot p_i)^2}
     {T_0\sqrt{(p_i\cdot q)^2-Q^2(p_i\cdot t)^2}}.
\end{align}
The first two lines in eq.~(\ref{eq:bw}) represent the variation in
$\Bw$ at fixed thrust axis, and the last line is the contribution due
to the fact that if the emission is in the hemisphere of the hardest
parton, the thrust axis is tilted, and this affects $\Bw$.  Notice
also that while the first term requires a subtraction (the two
following terms), the last term does not.  In fact, $l$ cannot be
collinear with the partons opposite to the hardest one. Thus, assuming
for example that parton $p_1$ is in the hemisphere opposite to the
hardest parton, there will be a cut-off for large values of
$\eta$. Therefore, large values of $\eta$ will only be allowed if $l$
is collinear to the hardest parton, i.e. the one aligned with the
thrust axis. In this case however, it is easy to check that the
numerator in the last line of eq.~(\ref{eq:bw}) vanishes.

For the calculation of $y_3$, we assume first that the two soft partons
arising from gluon
splitting are not the first pair to be clustered together.
Under this assumption, also $y_3$ becomes additive
in the soft partons, and the calculation can be done in analogy with the other
variables. Assume for definiteness that, at the Born level, the hard partons
pair yielding the smallest $y_3$ is given by the parton labels
$j,\,k$, that the remaining parton is labeled $i$, and that $p_k^0<p_j^0$.
Then the change in $y_3$ is given by
\begin{align}
  h_{y_3}(\eta,\phi)=&\Bigg\{\theta(d_{k,l}< \min(d_{j,l},d_{i,l}))
                       2\Bigg[  2 p_k^0 l^0(1-\cos \psi_{kj})
   - (p_k^0)^2\left(\frac{\vec{l}\cdot \vec{p}_j}{|\vec{p}_j|
                       |\vec{p}_k|} 
    - \frac{\vec{p}_j\cdot \vec{p}_k\, \vec{p}_k\cdot \vec{l} }{|\vec{p}_j| |\vec{p}_k|^3}\right)\Bigg]
     \nonumber \\
  & +\theta(d_{j,l}< \min(d_{k,l},d_{i,l}))
                       2\Bigg[ 
   - (p_k^0)^2\left(\frac{\vec{l}\cdot \vec{p}_k}{|\vec{p}_k|
                       |\vec{p}_j|} 
    - \frac{\vec{p}_j\cdot \vec{p}_k\, \vec{p}_j\cdot \vec{l} }{|\vec{p}_k| |\vec{p}_j|^3}\right)\Bigg] \nonumber \\
  & -\theta(d_{k,l^+}< \min(d_{j,l^+},d_{i,l^+})) 2(2p_k^0 (l^+)^0)(1-\cos \psi_{kj})\nonumber \\
  &
     -\theta(d_{k,l^-}< \min(d_{j,l^-},d_{i,l^-})) 2(2p_k^0 (l^-)^0)(1-\cos \psi_{kj}) \Bigg\}   \label{eq:hy3}                 
\end{align}
where
\begin{equation}
  d_{h,l}=1-\frac{\vec{p}_h\cdot \vec{l}}{|\vec{p}_h| |\vec{l}|},\quad\quad
  \mbox{and} \quad \cos \psi_{kj}=
  \frac{\vec{p}_k\cdot \vec{p}_j}{|\vec{p}_k| |\vec{p}_j|}.
\end{equation}
The term proportional to $p_k^0 l^0$ is due the change in the energy
of parton $k$ when it combines with $l$, while the terms proportional
to $(p_k^0)^2$ are due to the change in the angle between parton $k$
combined with $l$ and parton $j$ (the first instance), and between
parton $j$ combined with $l$ and parton $k$ (second instance).  Notice
that there are no subtractions associated with the change in angle.
In fact, because of the theta functions, the only collinear
singularity that can arise in this case is when $l$ is collinear to
$j$ or $k$, but then the angle does not change.

As stated earlier, the $y_3$ variable is really not additive, i.e. the
$y_3$ modification due to several soft emissions is not the sum of the
modifications due to each emission since partons can be clustered
together.\footnote{This is also discussed in
  ref.~\cite{Dasgupta:2009tm}. We thank Andrea Banfi for pointing this
  out to us.}
In order to estimate the magnitude of the error
associated with this assumption, it is interesting to compute the
non-perturbative correction to $y_3$ also in the case when the two
partons are always clustered together. In this case
formula~(\ref{eq:hy3}) still holds, with $l$ equal to the total
momentum of the pair of partons, $l^2=\lambda^2$, and, in the left
hand side, $h_{y_3}(\eta,\phi)$ is replaced by $h_{y_3}(l)$.  The
non-perturbative correction can then be written as
\begin{equation}\label{eq:npcorrnosplity3}
  \left[\Sigma(v)\right]_{\rm NP}=\left\{\int \mathd \sigma_B(\Phi_B) \delta(v(\Phi_B) -v)
  \sum_{\rm dip}\left[ 4 \frac{\as C_{\rm dip}}{2\pi}\frac{1}{Q}
    \int \mathd y\,\frac{\mathd \phi}{4\pi}
    \frac{\mathd l_\perp^2}{l_\perp^2+\lambda^2}h_{y_3}(l)\right]\right\}_\lambda
   \times I_{\rm NP},
\end{equation}
 where the suffix $\lambda$ in the closing curly bracket indicates
 that we should extract the coefficient of the term proportional to
 $\lambda$ in the enclosed expression. We will use
 formula~(\ref{eq:npcorrnosplity3}) in the following to assess the
 error due to our approximation in eq.~(\ref{eq:hy3}).

 \subsection{The shift in the cumulative cross section}
 In eq.~(\ref{eq:npcorr}) we have given the formula for the
 non-perturbative correction to the leading order 3-jet cross section.
 It is customary to express the non-perturbative correction as a shift
 in $\Sigma(v)$, i.e. to write
\begin{equation}\label{eq:cumulant}
  \Sigma_{\rm B+NP}(v)-\SigmaLO(v)=\SigmaLO\left(v-\delta v\right)-\SigmaLO(v)=-\frac{\mathd \sigma_B}{\mathd v}\delta v,
  \quad\quad \delta v=\Hnp \zeta(v),
\end{equation}
where $\SigmaLO(v)$ is the Born level value
\begin{equation}
  \SigmaLO(v)=\int_0^v \frac{\mathd \sigma_B}{\mathd v}\mathd v,
\end{equation}
and using eq.~(\ref{eq:npcorr}) for the left-hand side of eq.~(\ref{eq:cumulant}), we thus define
\begin{align}\label{eq:zetadef}
\zeta(v)&=\left(\frac{\mathd \sigma_B}{\mathd v}\right)^{-1}\left\{\int \mathd \sigma_B(\Phi_B) \delta(v(\Phi_B) -v)
  \left[\sum_{\rm dip} \frac{C_{\rm dip}}{\Cf}
             \int \mathd \eta\,\frac{\mathd \phi}{2\pi} h_v(\eta, \phi)\right]\right\}, \\
  \Hnp&={\cal M} \times 4 \frac{\as \Cf}{2\pi}\times \frac{I_{\rm NP}}{Q}.\label{eq:Hdef}
\end{align}
With the above normalization, the shift function $\zeta$ in the
two-jet case assumes the values $3\pi$ for the $C$ parameter, 2 for
$\tau=1-T$, 1 for $\Mh^2$ and 0 for $\Md^2$ and $y_3$.  For the wide
jet broadening in the two-jet limit the linear $\lambda$ term is
actually accompanied by a $\log \lambda/Q$, and thus the linear term
does not have a finite coefficient.

\begin{figure}[htb]
  \begin{center}
    \includegraphics[width=0.47\linewidth, page=1]{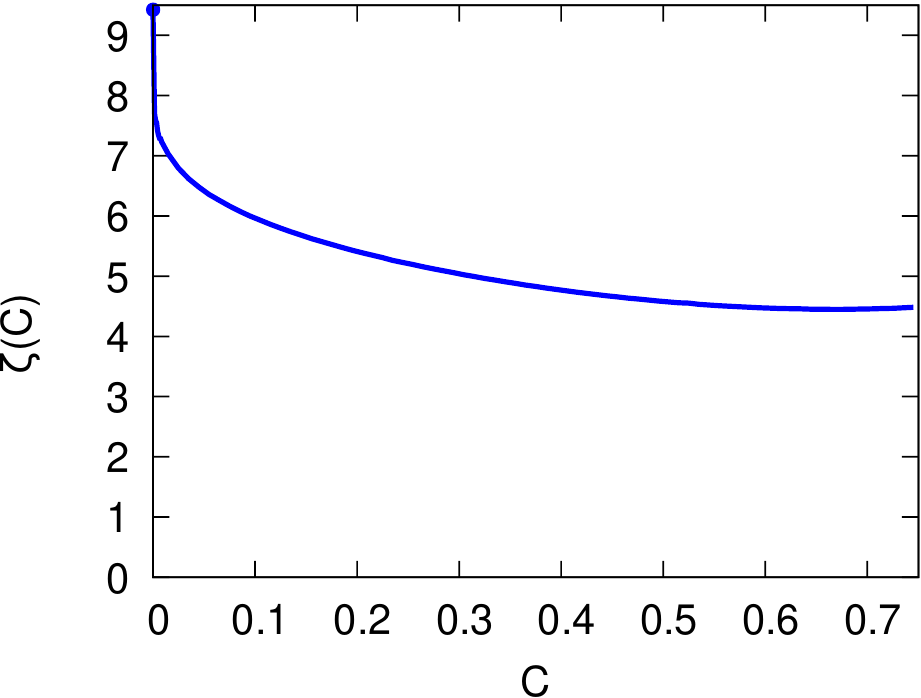}
    \hskip 0.04\linewidth
    \includegraphics[width=0.47\linewidth, page=2]{Figures/plotzeta}
    \vskip 0.03\linewidth
    \includegraphics[width=0.47\linewidth, page=3]{Figures/plotzeta}
    \hskip 0.04\linewidth
    \includegraphics[width=0.47\linewidth, page=4]{Figures/plotzeta}
    \vskip 0.03\linewidth
    \includegraphics[width=0.47\linewidth, page=5]{Figures/plotzeta}
    \hskip 0.04\linewidth
    \includegraphics[width=0.47\linewidth, page=6]{Figures/plotzeta}
    \caption{The function $\zeta$ plotted for $C$, $1-T$, $y_3$, $\Mh^2$, $\Md^2$ and $B_W$.\label{fig:zeta}}
  \end{center}
\end{figure}

\begin{figure}
   \begin{center}
   \includegraphics[width=0.6\textwidth]{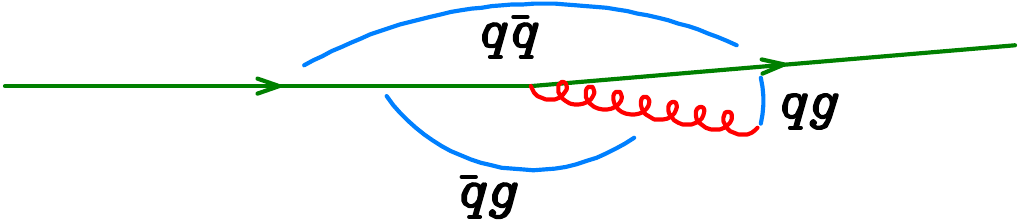}
   \caption{Dominant double logarithmic region near the two jet limit.
     The $qg$ dipole does not radiate, while the $q\bar{q}$ and $\bar{q} g$ dipoles differ
     only by their colour factor.\label{fig:softcoll}}
   \end{center}
 \end{figure}

We computed the functions $\zeta(v)$ for the variables listed above. The results are displayed in Fig.~\ref{fig:zeta}.
  With an angular-ordering argument, one can show that the limit
  $\zeta(v)$ for $v\to 0$ should tend to the corresponding two-jet
  limit values. In fact, in this limit, the emitted hard gluon becomes
  collinear to either the quark or the antiquark, let us say to the
  quark for sake of discussion, as shown in Fig.~\ref{fig:softcoll}.
 Because of coherence, the soft gluon associated with the power
 corrections sees the collinear quark-gluon pair as a single colour
 source, with the same colour of $q$.  Thus the emission pattern is
 the same as that of a quark-antiquark pair. Alternatively, one may
 consider the emissions of the three dipoles $q\bar{q}$, $qg$, and
 $\bar{q}g$, that carry the colour factors $\Cf-\Ca/2$ for $q\bar{q}$
 and $\Ca/2$ for $qg$ and $\bar{q}g$.  The $qg$ dipole does not emit
 in the small angle limit (the eikonal formula vanishes there), and
 the $\bar{q}g$ dipole becomes equal to the $q\bar{q}$ dipole, giving
 $\Cf-\Ca/2+\Ca/2=\Cf$, i.e. the same soft radiation of a $q\bar{q}$
 dipole.  This must happen, however, when the logarithm of the shape
 variable is so large that it clearly prevails over single logs and
 constant terms.  In the case of $C$ and $1-T$, one finds that for values of the
 shape variable $v\approx 10^{-3}$ the $\zeta$ function differs from
 the two-jet limit value by roughly 10\%{}, i.e. of the order of
 $1/\log(v)$, that is the natural size of single-log corrections.

\begin{figure}
   \begin{center}
   \includegraphics[height=0.35\textwidth,page=1]{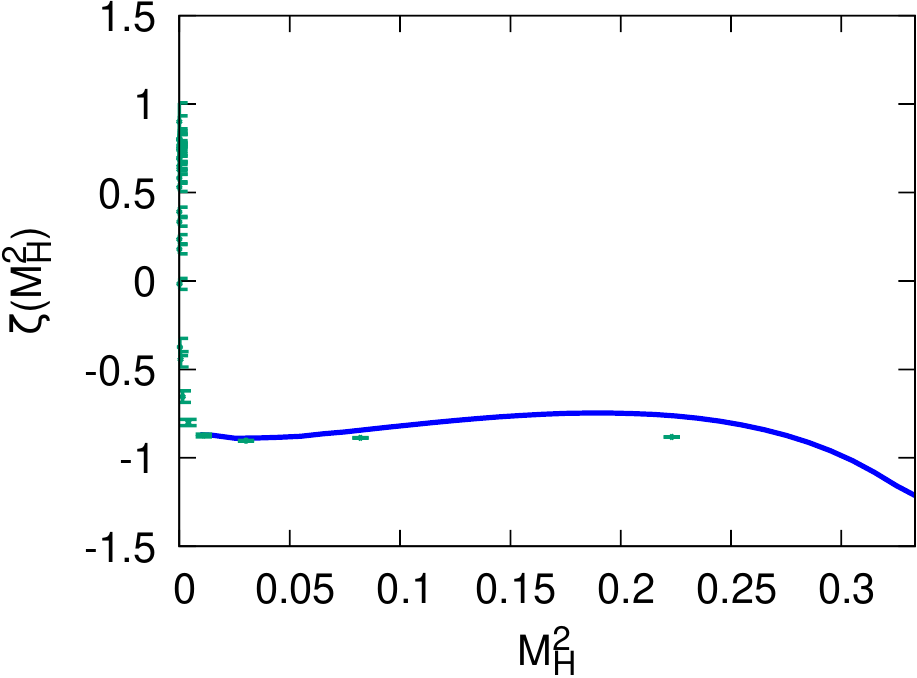}
   \includegraphics[height=0.35\textwidth,page=2]{Figures/plotmh2}
   \caption{The $\zeta(\Mh^2)$ function at very small value of its argument.
     The dots are obtained by performing a quadruple precision calculation and binning
     the results uniformly in a logarithmic scale. The left/right
     plot use a linear/logarithmic scale for the $x$ axis.\label{fig:twojetlimMh2}}
 \end{center}
 \end{figure}

 The case of $\Mh^2$ and $\Md^2$, however, are much more extreme.  In
 this case, in order to check that the two jet limit of $1$ and $0$
 respectively are actually reached, we had to perform a dedicated
 calculation in quadruple precision in the small $v$ region. As an
 example, we show in Fig.~\ref{fig:twojetlimMh2} the result of this
 calculation for $\Mh^2$.
 It is evident that $\Mh^2$ changes sign and reaches the value 1 very
 near zero, varying by about 2 units in a very narrow neighbourhood
 around zero. $\Md^2$ undergoes an even stronger variation, changing
 by three units, and reaching zero from negative values.  Such an
 abrupt change in the three-jet distribution as we approach
 the two-jet limit suggests that subleading soft terms in the two-jet
 limit remain more important than double logarithms all the way down
 to very small values of the shape variable, questioning on one side
 the possibility to associate the two-jet limit non-perturbative
 correction to the resummation of soft radiation, and, on the other
 side, the application of our newly computed non perturbative
 correction as we approach the two-jet limit.

 \subsection{Numerical checks}\label{sec:method2}
 As a numerical check of the above calculations we also computed the
 $\zeta$ functions by directly generating the phase space comprising
 the three hard partons and the soft one, fixing its
 transverse momentum to a value $\lambda_0=Q_0/100$. More explicitly,
 we first generate the underlying Born momenta $p_i$, $i=1\ldots 3$,
 choose $\lambda_0=1$ GeV and $Q_0=100$ GeV, and construct the momentum of the
 radiated parton as in eqs.~(\ref{eq:ldef}) to (\ref{eq:ldefprops}).
 Assuming for sake of argument that $p_1$ and $p_2$ are the momenta of
 the radiating dipole, we construct the recoil-corrected momenta as
 \begin{align}
   P_1&=p_1-l^+-\frac{1}{2} l_\perp\;, \nonumber \\
   P_2&=p_2-l^--\frac{1}{2} l_\perp\;.
 \end{align}
 In this way the total momentum is conserved, and the on-shell
 property of $P_{1/2}$ are maintained up to terms of order
 $\lambda^2/Q^2=1/10^4$.  The event comprising $P_1$, $P_2$, $p_3$
 and $l$ is then used to compute directly the values of the shape
 variables, and its difference with respect to the value obtained for
 momenta $p_1$, $p_2$ and $p_3$ is computed. Using this method, we
 find good agreement with the $\lambda\to 0$ calculations described in
 the previous section, except near the zero value of the shape
 variable and, in the case of the $C$ parameter, near the upper
 end-point of $3/4$, i.e. the 3-jet symmetric limit. We will make use
 of this method to give an estimate of corrections suppressed by
 higher powers of $\lambda$, as illustrated later.

\section{Calculation of the observable distributions}\label{sec:obscalc}

We are interested in fitting $\alpha_s$ from event shapes in the
three-jet region, where the novel results for the non-perturbative
corrections can be used. Furthermore, in the three-jet region the
relation between the observables and the value of $\alpha_s$ is more
direct.  For this reason, at the perturbative level we consider here only
fixed-order predictions and, when determining the fit range, we will
make sure that all-order resummed predictions, not included here, have
a small effect.

Perturbative predictions for $e^+e^-\to 3$~jets are available up to
next-to-next-to-leading order (NNLO) accuracy and are implemented in
the public code
EERAD3~\cite{Gehrmann-DeRidder:2007foh,Gehrmann-DeRidder:2007vsv,Gehrmann-DeRidder:2008qsl},
which is based on the antenna subtraction
formalism~\cite{Gehrmann-DeRidder:2005btv} and in a private
code~\cite{DelDuca:2016csb}, which is based on the CoLoRFulNNLO
subtraction method~\cite{DelDuca:2016ily}.
We have used here predictions from EERAD3 up to NNLO and have checked
that they agree with predictions using the CoLoRFulNNLO subtraction
method up to NLO accuracy.\footnote{We thank Adam Kardos for
  providing results up to NLO using the CoLoRFulNNLO subtraction
  method.}

Denoting by $v$ a generic event shape, the normalized integrated
distribution at center-of-mass energy $Q$ and at the renormalization
scale $\mu_R$ can be written as
\begin{eqnarray}
\Sigma_{\rm NNLO}(v) &=& \int_0^v dv' \frac{1}{\sigma_{\rm \scriptscriptstyle NNLO}}\frac{d\sigma_{\rm \scriptscriptstyle NNLO}(v',Q)}{d v'} = 
  \frac{\alpha_s(\muR)}{2\pi} \frac{dA(v)}{dv} \nonumber \\ &+&
  \left(\frac{\alpha_s(\muR)}{2\pi}\right)^2 \frac{dB(v,x_{\muR})}{dv} +
  \left(\frac{\alpha_s(\muR)}{2\pi}\right)^3 \frac{dC(v,x_{\muR})}{dv}\,, 
\end{eqnarray}  
where $\xR= \muR/Q$ and
\begin{eqnarray}
  B(v,x_{\muR}) &=& B(v,1) + A(v)\left(\beta_0 \ln\xR -\sigma_1 \right)\,,\\
  C(v,x_{\muR}) &=& C(v,1) + B(v,1)\left(2 \beta_0 \ln\xR-\sigma_1\right) \nonumber \\
  &+&A(v)\left(\frac12\beta_1\ln \xR +\beta_0^2  \ln^2\xR +\sigma_1^2-\sigma_2 \right)\nonumber\,,   
\end{eqnarray}  
with $\beta_0 = (11\Ca - 4 n_f T_R)/3$, $\beta_1 = (34\Ca^2 - 20 \Ca n_f T_R - 12 \Cf T_R n_f)/3$,
and where the expansion of the total cross section reads 
\begin{equation}
\sigma_{\rm NNLO} = \sigma_0 \left(1+\frac{\alpha_s(\muR)}{2\pi}\sigma_1 + \left(\frac{\alpha_s(\muR)}{2\pi}\right)^2\sigma_2\right)\,,  
\end{equation}  
with  $\sigma_1 = 3\Cf/2 $ and $\sigma_2 = \Cf\left((123/8-11\zeta_3)\Ca - 3/8 \Cf + (4 \zeta_3 - 11/2)\,n_f\, T_R\right)$. 
For our central predictions we choose $\xR = 1/2$, and we estimate the
error due to missing higher-order terms by varying this scale up and
down by a factor of two. The choice of $\xR = 1/2$ for the central
value is motivated by the fact that the scale entering in the
production of the third jet is somewhat lower than $Q$.

The non-perturbative corrections discussed in Sec.~\ref{sec:powcor}
can be included as a shift in the argument of the cumulative cross section,
i.e. according to eq.~(\ref{eq:cumulant}), but using instead the full
NNLO cross section. We now depart from the large-$\nf$
parameterization of the shift, and switch instead to the dispersive
model of ref.~\cite{Dokshitzer:1998pt}, where the role of the
effective coupling of eq.~(\ref{eq:largenfINP}) is played by a
parameter $\alpha_0(\mu_I^2)$. So, rather than using the definition of
eqs.~(\ref{eq:zetadef}) and (\ref{eq:Hdef}), the shift (see
eq.~(\ref{eq:cumulant})) can be written as
\begin{eqnarray}
  \delta v &=& \zeta(v) {\cal M} \frac{\mu_I}{Q} \frac{4 C_{\rm F}}{\pi^2}\Bigg[\alpha_0(\mu_I^2) -\alpha_s(\mu_R^2) 
     -  \alpha_s^2(\mu_R^2)\frac{\beta_0}{\pi}\left(2 \ln\frac{\mu_R}{\mu_I} +\frac{K^{(1)}}{2\beta_0} +2\right)\\ 
    & - & \left.\alpha_s^3(\mu_R^2)\frac{\beta_0^2}{\pi^2}\left(4 \ln^2\frac{\mu_R}{\mu_I} +4 \left(\ln\frac{\mu_R}{\mu_I}+1\right)\right) \times \left(2+\frac{\beta_1}{2\beta_0^2}+\frac{K^{(1)}}{2\beta_0}\right)+\frac{K^{(2)}}{4\beta_0^2}\right]\,, 
\label{eq:npshift}
\end{eqnarray}  
where the observable dependent part $\zeta(v)$ has been discussed in
detail in Sec.~\ref{sec:powcor}, the Milan factor is given in eq.~(\ref{eq:mifacnumeric})
and $\alpha_0(\mu_I^2)$ is defined
as a mean value of the strong coupling in the CMW~\cite{Catani:1990rr}
scheme below an infrared scale $\mu_I$ which is conventionally taken equal
to 2 GeV:
\begin{equation}
  \alpha_0(\mu_I^2) = \frac{1}{\mu_I} \int_0^{\mu_I} d\mu\, \tilde \alpha_s(\mu^2)\,, 
\end{equation}
where in the perturbative region the $\overline{\rm MS}$ and CMW couplings are related as
\begin{equation} \label{eq:tildealpha}
  \tilde \alpha_s(\mu^2) = \alpha_s(\mu^2) \left(1  + \frac{\alpha_s(\mu^2)}{2\pi} K^{(1)} + \left(\frac{\alpha_s(\mu^2)}{2\pi}\right)^2 K^{(2)} +  {\cal O}(\alpha_s^3)\right)\,,  
\end{equation}
with~\cite{Banfi:2018mcq,Catani:2019rvy}
\begin{eqnarray}
  K^{(1)} &=& \Ca\left(\frac{67}{18}-\frac{\pi^2}{6} \right) - \frac59 n_f \,, \label{eq:kappauno}\\
  K^{(2)} &=& \Ca^2\left(\frac{245}{24}-\frac{67}{9}\zeta_2 +\frac{11}{6}\zeta_3+\frac{11}{5}\zeta_2^2 \right) +\Cf n_f \left(-\frac{55}{24}+2\zeta_3\right)\\
  &+& \Ca n_f \left(-\frac{209}{108}+\frac{10}{9}\zeta_2-\frac{7}{3}\zeta_3\right)-\frac{1}{27}n_f^2
  +\frac{\beta_0}{2}\left(\Ca\left(\frac{808}{27}-28\zeta_3\right)-\frac{224}{54}n_f \right)\nonumber\,.
\end{eqnarray}
The last terms in Eq.~(\ref{eq:npshift}) are subtraction terms of
contributions already accounted for in the perturbative
calculation. This assumes that non-inclusive corrections are described
by the same multiplicative Milan factor ${\cal M}$, that applies to
all observables we consider with the exception of $y_3$, as discussed
at the end of section \ref{sec:other}.

Notice that in the large $\nf$ limit we found (see eq.~(\ref{eq:Hdef}))
\begin{equation}\label{eq:deltalargenf}
  \delta v =H_{\rm NP} \zeta(v)={\cal M} \times 4 \frac{\as \Cf}{2\pi}\zeta(v) \times \frac{I_{\rm NP}}{Q},
\end{equation}
where $I_{\rm NP}$ can be interpreted as the integral of the large-$\nf$, CMW effective coupling.
In fact, expanding the second line of eq.~(\ref{eq:largenfINP}) for small $\as$ we find
\begin{equation}
  I_{\rm NP}=\frac{1}{\as(\mu)}\int_0^{\mu_C} \mathd \lambda\, \as(\lambda e^{-5/6}),
\end{equation}
and
\begin{equation}
  \as(\mu e^{-5/6})
  \approx \as(\mu) + \frac{5}{3} b_{0,\nf}\as(\mu)^2  =  \as(\mu) \left(1-\frac{5}{9} \nf \frac{\as(\mu)}{2\pi}\right),
\end{equation}
consistently with eqs.~(\ref{eq:tildealpha}) and~(\ref{eq:kappauno}).

However, formula~(\ref{eq:deltalargenf}) differs by a factor $\pi/2$ with respect to eq.~(\ref{eq:npshift}),
i.e. the factor
$\Cf/(2\pi)$ is replaced by $\Cf/\pi^2$ in  eq.~(\ref{eq:npshift}). This replacement
(for more details see ref.~\cite{Dokshitzer:1998pt} near formula~(6.3)) is irrelevant for
the purposes of this work, but we follow this prescription in order to
fit values of $\alpha_0$ that can be compared
to those found in previous publications.

It is possible to implement the
non-perturbative corrections in different ways, leading to results that differ
by terms of order ${\cal O}(\alpha_s/Q)$. We use this ambiguity to assign an
uncertainty related to our treatment of non-perturbative
corrections. For this purpose we define four schemes. Our default
predictions, scheme ``(a)'', are obtained by shifting the
perturbative distribution $\Sigma_{\rm NNLO}(v)$ by the non-perturbative correction computed in
Sec.~\ref{sec:powcor}
\begin{eqnarray}
  \Sigma^{(a)} (v) = \Sigma_{\rm NNLO}(v-\delta v )\,.
  \label{eq:schemeA}
\end{eqnarray}
Furthermore, in scheme (a), we also add to $\delta v$ an approximate
estimate of quadratic corrections. These are obtained from the
difference between the numerical evaluation at finite transverse
momentum described in Sec.~\ref{sec:method2} with respect to our standard
calculation. More specifically, calling $\tilde\zeta(v)$ the evaluation
of Sec.~\ref{sec:method2}, we correct $\delta(v)$ as follows
\begin{equation}\label{eq:secondorder}
\delta(v)=\zeta(v)  H_{\rm NP} +
  \left(\tilde\zeta(v)\times \frac{Q_0}{\lambda_0}-\zeta(v)\right)\times \frac{Q_0}{\lambda_0}
\times H_{\rm NP}^2\,.
\end{equation}
Alternatively, instead of shifting the full NNLO distribution,
one can shift only in the leading order term $\SigmaLO$ of the
integrated distribution (scheme (b)):
\begin{eqnarray}
  \SigmaFULL^{(b)} (v) = \SigmaLO (v-\delta v) +\Sigma (v) -\SigmaLO (v)\,.
  \label{eq:schemeB}
\end{eqnarray}  
Yet another option is to expand the integrated distribution around the perturbative value (scheme (c)):
\begin{eqnarray}
  \SigmaFULL^{(c)} (v) = \Sigma (v) -\delta v \frac{\SigmaLO (v)}{dv}\,. 
  \label{eq:schemeC}
\end{eqnarray}
Scheme (d) is defined as scheme (a) but without the quadratic
correction of eq.~\eqref{eq:secondorder} included in the other schemes. 

\section{Fit to ALEPH data}\label{sec:fitALEPH}
We now compare the theoretical predictions including power corrections
to the ALEPH data of ref.~\cite{ALEPH:2003obs}, where several
shape variables were analyzed in the centre-of-mass energy range from
91.2 to 206~GeV. Here we focus upon the 91.2 GeV data.  Including
higher energy data does not lead to noticeable differences in the results,
as we will discuss briefly in Sec.~\ref{sec:he}.

Our goal is to fit several observables at once. We need to select
observables such that the power corrections in the three jet region
can be computed with our methods, and that are at the same time
available in ALEPH. These are the $C$-parameter, $\tau=1-T$, $y_3$
in the Durham scheme, the heavy-jet mass $\Mh^2$, the mass
difference $\Md^2$, and the wide jet broadening $B_W$.
Since the $y_3$ variable is not really additive, we need to provide an
estimate of the error associated with this. We will do so along the
lines discussed at the end of Sec.~\ref{sec:other}.

The non-perturbative corrections to $\Mh^2$, $\Md^2$ and $B_W$ have a
common feature: in the 3-jet regime they differ drastically from their
value in the two-jet limit. Such an abrupt change is quite worrisome,
and may be taken as an indication that higher-order emissions may be
associated with large corrections to the non-perturbative
coefficient. For this reason, initially we leave these variables out
of the fit, and only fit the $C$-parameter, $\tau$ and $y_3$.  We fit
the value of $\as(\Mz)$ and the non-perturbative parameter $\alpha_0$,
defined in Sec.~\ref{sec:obscalc}.

\subsection{Treatment of uncertainties}
\subsubsection{Statistical and systematic errors, and correlations}
\label{sec:err}
The ALEPH data (available at the site
\url{https://www.hepdata.net/record/ins636645}) includes statistical
and systematic errors.  Our method of choice for computing the error
is the following.  Calling $R_i$ the statistical error, $S_i$ the
systematic error, $T_i$ the theoretical error relative to bin $i$,
$C_{ij}$ the statistical correlation matrix, and ${\rm Cov}^{(\rm
  Sys)}_{ij}$ the covariance matrix for the systematic errors, we
compute the full covariance matrix as
\begin{equation}
  V_{i j}=\delta_{ij}(R_i^2+T_i^2) + (1-\delta_{ij})C_{ij}R_iR_j + {\rm Cov}^{(\rm Sys)}_{ij} \,,
\end{equation}
where the indices $i$ and $j$ run over all the bins of all observables
that have been included in the fit.  The ALEPH data  quotes two kinds
of systematic errors for the data taken at the $Z$ pole. We add these
two errors in quadrature to obtain the global systematic error that we
use in our analysis.

We computed the statistical correlation coefficients $C_{ij}$ using
\Pythiaeight{}.  Calling $N_i$ and $N_j$ the number of events that
fall into bin $i$ and bin $j$, and $N_{ij}$ the number of events that
contribute to both bins, we have
\begin{equation}
  C_{ij}=\frac{\frac{N_{ij}}{N}-\frac{N_i N_j}{N^2}}{\sqrt{\frac{N_i}{N}-\frac{N_i^2}{N^2}}\sqrt{\frac{N_j}{N}-\frac{N_j^2}{N^2}}},
\end{equation}
where $N$ is the total number of events.  Note that $N_{ij}$ is zero
for different bins of the same observable, so that a negative
correlation is expected for all pairs of bins in this case.

Statistical, systematic and theoretical errors are assumed to be
uncorrelated among each other.  For the covariance of the systematic
errors we adopt the so called ``minimum overlap'' assumption (denoted
in the following as MO), and set them equal to the minimum of the
square of the systematic errors for the bins in question, i.e.
\begin{equation}
  {\rm Cov}^{(\rm Sys)}_{ij} =  \delta_{ij}S_i^2+(1-\delta_{ij})\min(S_i^2, S_j^2)\,.
  \label{eq:MO}
\end{equation}
As an alternative, we computed the covariance matrix for the case of
$C$, $T$ and $y_3$, by using the 24 systematic variations of the
resulting distributions that were obtained by ALEPH in order to
determine the systematic errors.\footnote{We thank Hasko Stenzel for
  providing these data to us.}  We compute the covariance matrix and
the central value as follows. We call $v^{(r)}_i$ the value of a shape
variable for the bin $i$, where again $i$ denotes both the bin and the
observable, and where $r$ labels the 25 replicas (a central value plus
24 variations.). We then define
\begin{align}
  \bar{v}_i &=\frac{1}{N_r}\sum_r v^{(r)}_i, \\
  \bar{v}_{ij} &=\frac{1}{N_r}\sum_r v^{(r)}_i v^{(r)}_j, \\
  {\rm Cov}^{(\rm Sys)}_{ij} &= \sum_r \left(v^{(r)}_i-\bar{v}_i \right) \left(v^{(r)}_j-\bar{v}_j \right)
  = N_r \left(\bar{v}_{ij}-\bar{v}_i\bar{v}_j\right)\,.
  \label{eq:Rep}
\end{align}
We use  ${\rm Cov}^{(\rm Sys)}_{ij}$ as covariance matrix, and for the
central value we use either the replica corresponding to
the ALEPH default setup,
or the average over all replicas $\bar{v}_i$. Some of
the variations provided are double sided (i.e.\ they are associated
with a positive and negative variation of a parameter). For these variations we
have included a factor $1/2$ in the computation of $\bar{v}_{ij}$. In
the following we call this the ``replica method'', and denote it with R.

The covariance matrix is used to compute the $\chi^2$ value according
to the standard formula
\begin{equation}
  \chi^2=\sum_{ij} \left(v_i-v_i^{(\rm th)}\right) V_{ij} \left(v_j-v_j^{(\rm th)}\right).
\end{equation}

\subsubsection{Perturbative theory uncertainties}
As a consequence of the high precision of the LEPI data, in order to
obtain reasonable $\chi^2$ values when performing the fits we add the
theoretical uncertainty in quadrature to the experimental one.
We do not assume any correlations for the theoretical errors.

We define the perturbative theoretical error by considering three
values for the renormalization scale: $Q$, $Q/2$ and $Q/4$. Calling
$O(\mur)$ the value of a shape variable in a bin, we define the
perturbative central value $O_{\rm cv}$ and the associated
perturbative error $O_{\rm err}$ of the theoretical
prediction as follows,
\begin{align}
  O_{\rm cv}&=\frac{\max(O(Q), O(Q/2), O(Q/4))+\min(O(Q), O(Q/2), O(Q/4))}{2}, \\
  O_{\rm err}&=\frac{\max(O(Q), O(Q/2), O(Q/4))-\min(O(Q), O(Q/2), O(Q/4))}{2}.
\end{align}
The perturbative theoretical error is quite small at the NNLO level we
are working with.

\subsubsection{Non-perturbative theory uncertainty}
Non-perturbative corrections can be sizeable, up to the order of
10\%{}, and thus we must also include an error associated with them.
As seen in Sec.~\ref{sec:powcor}, there is evidence that power
suppressed corrections of second order are not negligible, especially
near the two jet region. 
We have estimated them and used them to correct the central value, see
Eq.~\eqref{eq:secondorder}. We thus associated an uncertainty equal to
twice the quadratic correction.
As a further point, we expect that the $\zeta$ functions may receive
perturbative corrections of order $\as\sim 0.1$.
We thus define the following associated error to $\delta(v)$ 
\begin{equation}\label{eq:nonperterr}
  \delta_{\rm err}(v)= 2\cdot\left|\tilde\zeta(v)\times
  \frac{Q_0}{\lambda_0}-\zeta(v)\right|\times
  \frac{Q_0}{\lambda_0}\times H_{\rm NP}^2+0.1 \cdot \delta(v)\,.
\end{equation}

\subsection{Correction for heavy-quark mass effects}
\label{sec:heavytolight}
\begin{figure}[htb]
  \begin{center}
    \includegraphics[width=0.8\linewidth]%
    {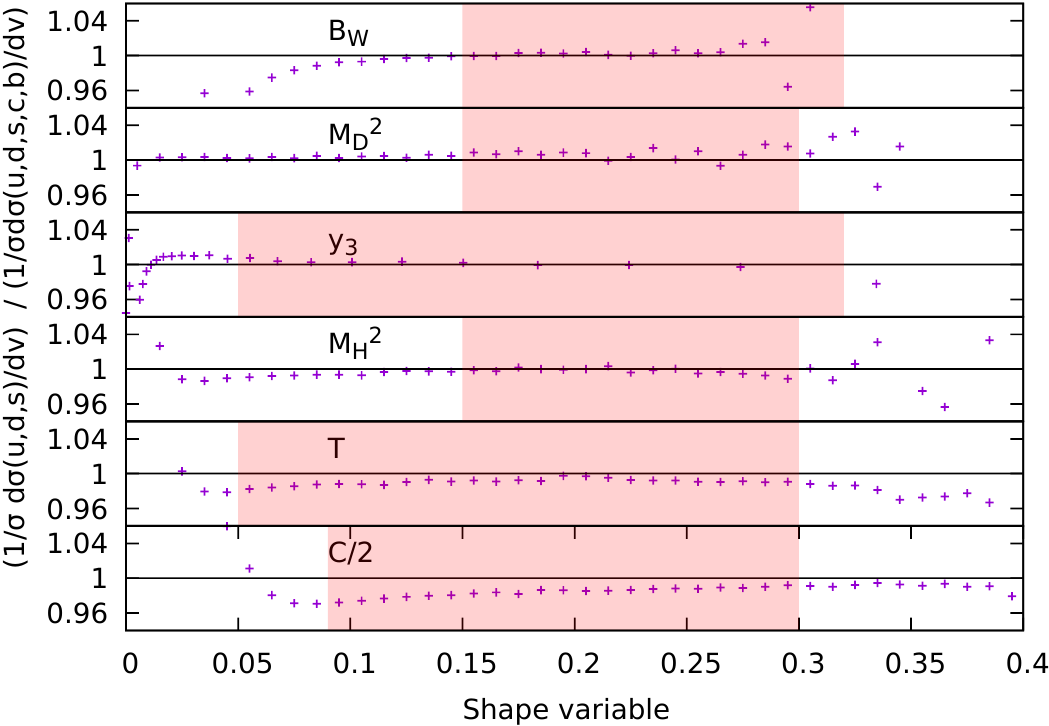}
    \caption{Heavy-flavours correction factors. The coloured band
      marks the range where fits are usually performed. Notice that
      we plot $C/2$ rather than $C$.\label{fig:heavyToLight}}
  \end{center}
 
\end{figure}

Our NNLO calculation deals with massless quarks, while the data
includes primary charm and bottom pairs.  We correct the data by
multiplying, for each bin of each observable denoted globally as $i$,
the ratio of the Monte Carlo predictions for the corresponding
observables $v_i$ evaluated without and with the $c$ and $b$ primary
production processes
\begin{equation}
  v_i^{\rm (corr)}= v_i\times \frac{v_i^{{\rm MC,} uds}}{v_i^{{\rm MC,} udscb}}.
\end{equation}
The correction factors obtained using \PythiaEight are shown in Fig.~\ref{fig:heavyToLight}.
Notice that corrections are quite modest, although not totally negligible in some cases.

\subsection{Hadron mass-effects corrections}
\label{sec:schemes}
As already discussed in Sec.~\ref{sec:masses}, the theoretical
calculation of shape-variable distributions deals with massless
particles and the massless definition can be extended to deal with
massive particles using different schemes.
Since full particle identification is not available in an experimental
context, this lack of information is filled by the Monte Carlo
simulation when correcting from the detector level to the generator level.
We also use a Monte Carlo generator to compute
shape variables in the different schemes, and then construct migration
matrices to correct from the scheme adopted by the experiment to any
another scheme. More specifically, for each Monte Carlo event, we
compute the shape variable in the standard scheme (the one adopted by
the experiment, as defined in Sec.~\ref{sec:shapevars})
and another scheme $S$. Assuming that the shape
variable in the standard scheme falls into bin $i$, and the same shape
variable in scheme $S$ falls into bin $j$, we increase by one unit a
migration matrix $T_{ij}$. This matrix is used to correct the real
data according to the formula
\begin{equation}
  n_j^{\scriptscriptstyle (S)} = \sum_i n^{\scriptscriptstyle \rm
    (data)}_i \frac{T_{ij}}{\sum_k T_{ik}},
\end{equation}
designed in such a way that if one replaces the $n^{\scriptscriptstyle
  (data)}_i$ with its Monte Carlo prediction, one obtains by
construction the Monte Carlo prediction for $n_j^{\scriptscriptstyle
  (S)}$.
In the following, we use this method to assess the hadron-mass
sensitivity of our results.

\section{Fit results}\label{sec:fit}

Our default fit is based on the ALEPH data of
ref.~\cite{ALEPH:2003obs} at 91.2 GeV, and includes the thrust
variable $\tau=1-T$, the $C$-parameter and the Durham 3-jet resolution
variable $y_{3}$.
In our perturbative predictions we fix the renormalization scale to
$\mu_R = Q/2$.
Non-perturbative effects are included as a shift of the total
integrated distribution, corresponding to scheme (a) in Eq.~\eqref{eq:schemeA}.
Our default mass scheme is the E scheme discussed in
Sec.~\ref{sec:masses}, since it yields intermediate results with
respect to the other schemes, and is also closer to the result
obtained in the standard scheme (i.e.\ the scheme used by ALEPH,
as defined in Sec.~\ref{sec:shapevars}).
The treatment of correlations is described in Sec.~\ref{sec:err}. In
particular, we chose the minimum-overlap method as our default choice,
see Eq.~\eqref{eq:MO}.
We apply the heavy-to-light correction factors described in
Sec.~\ref{sec:heavytolight} and illustrated in
Fig.~\ref{fig:heavyToLight}.
We use \PythiaEight as our standard Monte Carlo to compute the
heavy-to-light correction factor and, when using a different mass
scheme, to compute the migration matrix to be used to correct from
the scheme adopted by the experiment to any another scheme.
To perform our fit we use the default fit ranges listed in the second
column of Table~\ref{tab:ranges}.  
\begin{table}[htb]
\begin{center}
\begin{tabular}{|c|c|c|c|}
  \hline
  observable & default &  Fit ranges (2)&  Fit ranges (3) \\
\hline 
  $C$ & [\,0.25\,:\,0.6\,] & [\,0.17\,:\,0.6\,] & [\,0.375\,:\,0.6\,] \\
\hline
$\tau$ & [\,0.1\,:\,0.3\,] & [\,0.067\,:\,0.3\,] & [\,0.15\,:\,0.3\,] \\
\hline
$y_{3}$ & [\,0.05\,:\,0.3\,] & [\,0.033\,:\,0.3\,] & [\,0.075\,:\,0.3\,] \\
  \hline
\end{tabular}
\end{center}
\caption{Default fit range used (second column), and alternative
  choices (obtained by multiplying the default lower bound by 2/3 and 3/2)
  used to estimate the impact of the choice of the fit range
  (third and forth column).}
  \label{tab:ranges}
\end{table}
The lower edges of the ranges are determined in such a way that the
impact of the resummation remains small (see
Appendix~\ref{app:resum}), while the upper edge is close to the
three-particle kinematic bound of the observable.
The result of the simultaneous fit of
$\alpha_s$ and $\alpha_0$, together with the total $\chi^2$ and
$\chi^2$ per degree of freedom is shown in the first line of
Table~\ref{tab:fits}.
\begin{table}[htb]
\begin{center}
  \begin{tabular}{|c|c|c|c|c|}
  \hline Variation & $\alpha_s(M_Z)$ & $\alpha_0$ & $\chi^2$ &
  $\chi^2/N_{\rm deg}$ \\
  \hline 
{\bf Default setup}              & {\bf 0.1182} & {\bf 0.64} & {\bf 7.3} &{\bf 0.17} \\
\hline
Renormalization scale $Q/4$      & 0.1202 & 0.60 &  9.1 & 0.21\\
Renormalization scale $Q$        & 0.1184 & 0.68 &  8.7 & 0.20\\
\hline
NP scheme (b)                    & 0.1198 & 0.77 &  7.0 & 0.16\\
NP scheme (c)                    & 0.1206 & 0.80 &  5.4 & 0.12\\
NP scheme (d)                    & 0.1194 & 0.66 &  5.8 & 0.13\\
\hline
$P$-scheme                       & 0.1158 & 0.62 & 10.7 & 0.24\\
$D$-scheme                       & 0.1198 & 0.79 &  5.7 & 0.13\\
Standard scheme                  & 0.1176 & 0.58 &  9.2 & 0.21\\
\hline
No heavy-to-light correction     & 0.1186 & 0.67 &  6.8 & 0.16\\
\hline
\HerwigSix{}                     & 0.1180 & 0.59 & 15.9 & 0.36\\
\HerwigSeven{}                   & 0.1180 & 0.60 & 12.0 & 0.27\\
\hline
Ranges (2)                       & 0.1174 & 0.62 & 12.7 & 0.23\\
Ranges (3)                       & 0.1188 & 0.69 &  2.7 & 0.08\\
\hline
Replica method (around average)  & 0.1192 & 0.61 &  7.0 & 0.16\\
Replica method (around default)  & 0.1192 & 0.61 &  7.0 & 0.16\\
\hline
$y_3$ clustered                  & 0.1174 & 0.66 &  8.2 & 0.19\\
\hline
$C$                              & 0.1256 & 0.48 &  1.3 & 0.07\\
$\tau$                           & 0.1194 & 0.64 &  0.8 & 0.04\\
$y_3$                            & 0.1214 & 1.81 &  0.2 & 0.02\\
$C$, $\tau$                      & 0.1238 & 0.51 &  2.6 & 0.07\\
\hline
  \end{tabular}
\end{center}
\caption{Default fit result for $\alpha_s(M_Z)$ and $\alpha_0$ (first line)
  and other fit results obtained by varying the setup. See text for
  more details.}
\label{tab:fits}
\end{table}

In the same table we illustrate how the fit results change if we vary
any of the default choice made.  In particular, we show the fit
results when fixing the central value of the renormalization scale to
$\mu_R=Q/4$ or $Q$.
We investigate the impact of the way in which non-perturbative
corrections are implemented, using the alternative schemes (b, c, d) presented in
Sec.~\ref{sec:obscalc} (near Eq.~\eqref{eq:schemeB}).
We also present the result obtained using the $P$- and $D$- scheme to define the
observables, as discussed in Sec.~\ref{sec:schemes}, and the result obtained
in the standard scheme.
To assess the impact of the heavy-to-light correction factor we switch
it completely off.
We vary the Monte Carlo used to compute the migration matrix for the scheme
and the heavy-to-light correction factor, and
consider \mbox{\HerwigSix\cite{Corcella:2000bw}} and \HerwigSeven\cite{Bellm:2015jjp}.
We vary the fit ranges adopted, as detailed in columns three and four of
Table~\ref{tab:ranges}.
Since correlations play an important role, we also use the replica
method, see Eq.~\eqref{eq:Rep}, 
using variations either around the average values of the replicas, or
around the values of the default replica.

In the case of $y_3$ there is one further uncertainty,
associated with
the fact that we computed the non-perturbative correction assuming
that the two soft partons from the splitting of the soft gluon are not
clustered together.  In order to estimate an associated uncertainty,
we also computed the non-perturbative correction assuming that the two
soft partons are always clustered together, see
Eq.~\eqref{eq:npcorrnosplity3}. The ratio of the latter to the former
results ranges from $0.7$ up to $0.85$ in the fit window adopted for
$y_3$. We have therefore performed the fit using alternatively the
approximation where the soft partons are always clustered together.
The corresponding result is reported in the table labeled as
$y_3$-clustered.  The central value for $\as$ in the simultaneous fit
of $y_3$, $C$ and $T$ is reduced by 0.7\%{}.  Given the fact that we
have chosen the lowest extreme of the variation, and that the correct
result must lie between the always-clustered and the never-clustered
cases, our estimate of this uncertainty is very
conservative.\footnote{Notice that the anti-$k_t$
  algorithms~\cite{Cacciari:2008gp} are such that the softest
  particles are never clustered together.}

Finally, we examine how the fit results change if we consider one
observable at the time, or if we exclude $y_3$ from the fits.
\subsection{Including higher energy data}\label{sec:he}
In the ALEPH publication~\cite{ALEPH:2003obs}, data are also available
for centre-of-mass energies of 133, 161, 172, 183, 189, 200 and
206~GeV. Including these data does not appreciably change the result of the
fit. For our default setup we get $\as(\Mz)=0.1184$ and $\alpha_0=0.64$,
compared to $\as(\Mz)=0.1182$ and $\alpha_0=0.64$ of the fit on the $Z$ peak.
We get a $\chi^2/N_{\rm deg}=0.70$, larger than the $0.17$ of the table. This
is easily understood, since higher energy data have dominant statistical errors,
and thus the $\chi^2/N_{\rm deg}$ is more in line with the expectation from statistical
dominated data.

\subsection{Discussion of the results}
Our findings can be summarized as follows. For all results presented
in the table, we observe an excellent $\chi^2$ of the fit. In
particular the $\chi^2$ over number of degrees of freedom is always
well below one. This is a consequence of our treatment of the
theoretical error, that has been added bin-by-bin to the experimental
one without correlations.
Because of this, the theoretical prediction has considerable
flexibility to adapt to data.

The choice of renormalization scale changes the fit by about 1.5\%,
the largest change driven by the variation to lower scales.
A similar change can be observed when examining alternative schemes to
implement non-perturbative corrections. 
The mass-scheme definitions bring in an effect of about 2\%.
The heavy-to-light correction factor changes $\alpha_s$ by just about
one permille, hence the uncertainty associated to this correction seems
negligible.
A few permille differences are found when using a different Monte
Carlo to change from the standard definition to the E-scheme and to
perform the heavy-to-light correction. These small differences are not
surprising since all the Monte Carlos we use are tuned to these data.
The choice of the fit range has an impact on the result of about one
percent. This confirms that the range chosen is such that the impact
of the resummation is modest. 
The choice of how to treat statistical correlations has also a similar
impact, and confirms that our minimal overlap approach provides a sensible
description of the correlations.
For $y_3$, the difference between the two limiting cases (where soft
emissions are always-clustered or never-clustered) amounts also to
about a one percent effect on the full fit.

\begin{figure}[htb]
  \begin{center}
    \includegraphics[width=0.7\linewidth]%
    {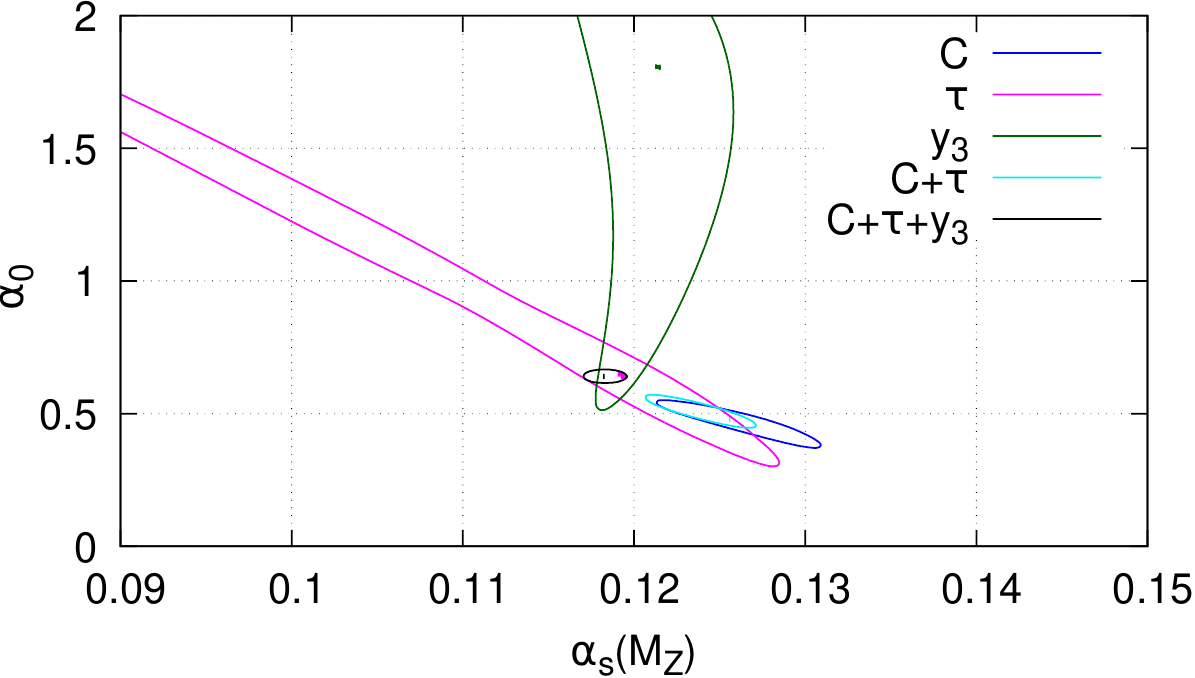}
    \caption{Contours at $\Delta\chi^2=1$ for fitting, $C$, $\tau$ and  $y_3$ individually,
      and then in the combinations $C+\tau$ and $C+\tau+y_3$.
     \label{fig:CTtau}}
  \end{center}
\end{figure}

Finally, we note that if one fits $\alpha_s$ and $\alpha_0$ from the
three observables considered separately, one tends to get a larger
value of the strong coupling, but with very different values of
$\alpha_0$. Indeed, there is a tension in the fitted value of
$\alpha_0$, where both thrust and $C$-parameter prefer a lower value,
while $y_{3}$ prefers a higher one. When fitting all observables at
the same time, the overall effect is that one finds an intermediate
value for $\alpha_0$ and a lower value of $\alpha_s$. The $\chi^2$ of
the fits remain excellent, which justifies a simultaneous fit. The
role of each variable in the common fit is illustrated in
Fig.~\ref{fig:CTtau}.
As one can see, for $C$ and $\tau$, $\alpha_0$ and $\alpha_S$ are
strongly anti-correlated, and with a similar anti-correlation. On the
other hand, $y_3$ has a $\zeta$ function that is small and of opposite sign,
and thus $\alpha_0$ and $\alpha_S$ are only weakly
correlated. The combined fit is then strongly constrained leading to
an intermediate value of $\alpha_0$ and a smaller value of $\as$.

Altogether, we conclude by remarking that our fit results agree very
well with the world average. In particular, we do not find low values
of $\alpha_s$ for the thrust or $C$-parameter which are included in
the current PDG average~\cite{ParticleDataGroup:2020ssz}.  However,
our results also clearly show that a fit of $\alpha_s$ from event
shapes with an overall uncertainty below the percent level seems today
not feasible. In particular, by changing certain choices that we have
made, like the central renormalization scale or the mass scheme, one
can easily obtain higher values of $\alpha_S$.

\subsection{Comparison to results obtained by setting $\zeta(v) = \zetatj(v)$}

It is natural now to ask what the results of the fits would have been
if we had used the non-perturbative correction as estimated in the
two-jet limit. For $C$, $\tau$, $y_3$, $\Mh^2$ and $\Md^2$ this
amounts to setting the $\zeta(v)$ functions plotted in
Fig.~\ref{fig:zeta} to a constant value, according to the
table~\ref{tab:zeta0}.\footnote{\label{foot:2jlim} For the case of $y_3$, the
  coefficient is known to be zero~\cite{Dokshitzer:1995qm}, since
  $y_3$ is quadratic in the transverse momentum for soft emissions. As
  for the case of $\Md^2$, a colour coherence argument would lead to
  the conclusion that in the dominant collinear limit the corrections
  to the two hemispheres are identical, leading again to a null value.
  For the remaining variables, see for example table~1 of
  ref.~\cite{Salam:2001bd}.}  For $\Bw$ the function $\zetatj(v)$ can
be found in Appendix~F of ref.~\cite{Dokshitzer:1998qp}.
\begin{table}[htb]
\begin{center}
\begin{tabular}{|c|c|c|c|c|c|c|}
  \hline
  $v$ & $C$ & $\tau$ & $y_3$ & $\Mh^2$ & $\Md^2$ & $B_W$ \\ 
  \hline
  $\zetatj(v)$ & $3\pi$ & $2$ & $0$ & $1$ & $0$ & App.~F of \cite{Dokshitzer:1998qp}  \\
  \hline
\end{tabular}
\caption{The non-perturbative coefficients in the two jet limit.}\label{tab:zeta0}
\end{center}
\end{table}
The complete results are reported in table~\ref{tab:fits-constzeta}.
\begin{table}
\begin{center}
  \begin{tabular}{|c|c|c|c|c|}
  \hline Variation & $\alpha_s(M_Z)$ & $\alpha_0$ & $\chi^2$ &
  $\chi^2/N_{\rm deg}$ \\
  \hline 
{\bf Default setup}              & 0.1132 & 0.55 & 15.8 & 0.36\\
\hline
Renormalization scale $Q/4$      & 0.1174 & 0.53 &  8.5 & 0.19\\
Renormalization scale $Q$        & 0.1126 & 0.57 & 22.0 & 0.50\\
\hline
NP scheme (b)                    & 0.1126 & 0.63 & 25.7 & 0.58\\
NP scheme (c)                    & 0.1134 & 0.72 & 16.4 & 0.37\\
NP scheme (d)                    & 0.1132 & 0.55 & 15.8 & 0.36\\
\hline
$P$-scheme                       & 0.1108 & 0.53 & 21.8 & 0.50\\
$D$-scheme                       & 0.1126 & 0.66 & 16.1 & 0.37\\
Standard scheme                  & 0.1134 & 0.51 & 15.9 & 0.36\\
\hline
No heavy-to-light correction     & 0.1130 & 0.58 & 15.9 & 0.36\\
\hline
\HerwigSix{}                     & 0.1136 & 0.51 & 31.1 & 0.71\\
\HerwigSeven{}                   & 0.1136 & 0.52 & 21.8 & 0.49\\
\hline
Ranges (2)                       & 0.1122 & 0.54 & 30.0 & 0.55\\
Ranges (3)                       & 0.1134 & 0.58 & 10.5 & 0.33\\
\hline
Replica method (around average)  & 0.1158 & 0.53 & 13.4 & 0.31\\
Replica method (around default)  & 0.1160 & 0.53 & 13.5 & 0.31\\
\hline
$y_3$ clustered                  & 0.1132 & 0.55 & 15.8 & 0.36\\
\hline
$C$                              & 0.1238 & 0.45 &  1.3 & 0.08\\
$\tau$                           & 0.1202 & 0.51 &  1.2 & 0.06\\
$y_3$                            & 0.1160 & -- &  1.4 & 0.18\\
$C$, $\tau$                      & 0.1222 & 0.46 &  2.7 & 0.08\\
\hline
  \end{tabular}
\end{center}
\caption{Default fit result for $\alpha_s(M_Z)$ and $\alpha_0$ (first line)
  and other fit results obtained by varying the setup, using the
  $\zetatj$ values of table~\ref{tab:zeta0}. See text for more
  details.}
\label{tab:fits-constzeta}
\end{table}
As shown there, the values of $\as$ found in this way are consistently lower
than those of table~\ref{tab:fits}. For example, for our default setup
we have $\as(\Mz)=0.1182$ and $\alpha_0=0.64$, while using
$\zetatj$ we get $0.1132$ and $0.55$ respectively. On the other hand,
the $\chi^2$ values are also quite acceptable.\footnote{We do not ascribe any
  significance to the larger $\chi^2$ values in the two-jet limit, because
  in this case in eq.~\eqref{eq:nonperterr}
  we have assumed rather arbitrarily $\zeta^{(2)}/\zeta=0.1$.}

\begin{figure}[htb]
  \begin{center}
    \includegraphics[width=0.9\linewidth]{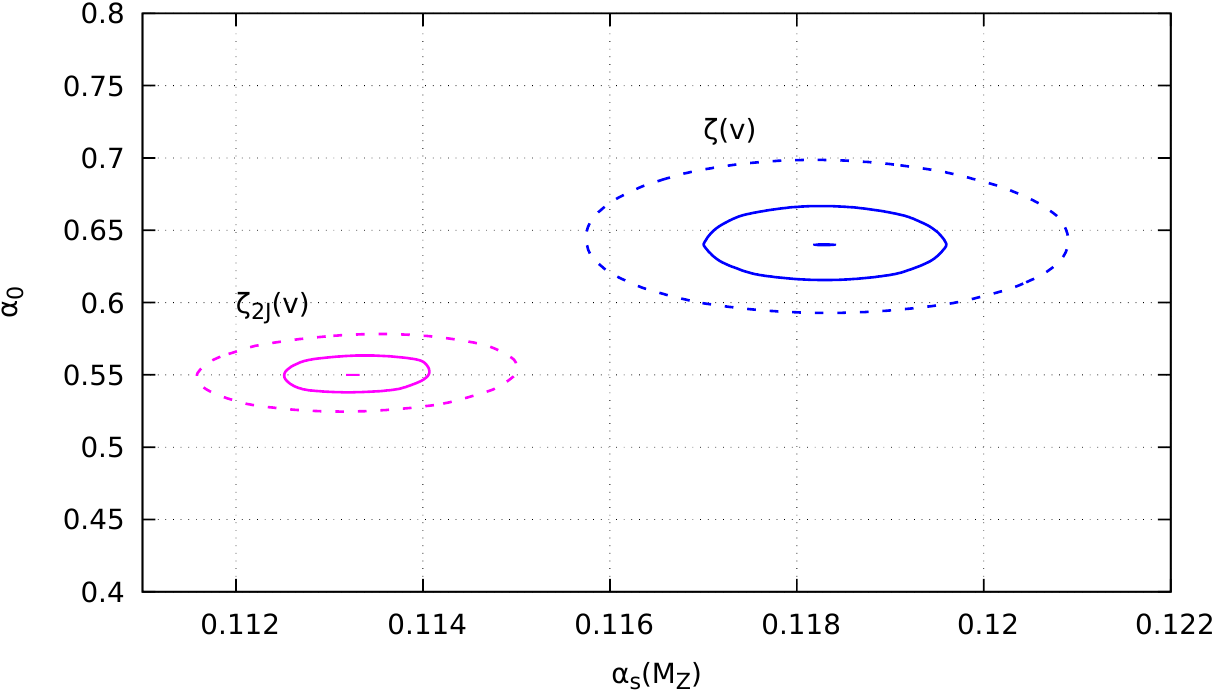}
    \caption{Central values and $\delta \chi^2=4$ (dashes) and 1 (solid) contours for our default
      fit of table~\ref{tab:fits} (blue) and the fit obtained with the $\zetatj$ functions,
      corresponding to the default fit of table~\ref{tab:fits-constzeta} (magenta).}
     \label{fig:as-a0-full-const}
  \end{center}
\end{figure}

\begin{figure}[htb]
  \begin{center}
    \includegraphics[width=0.495\textwidth,page=1]{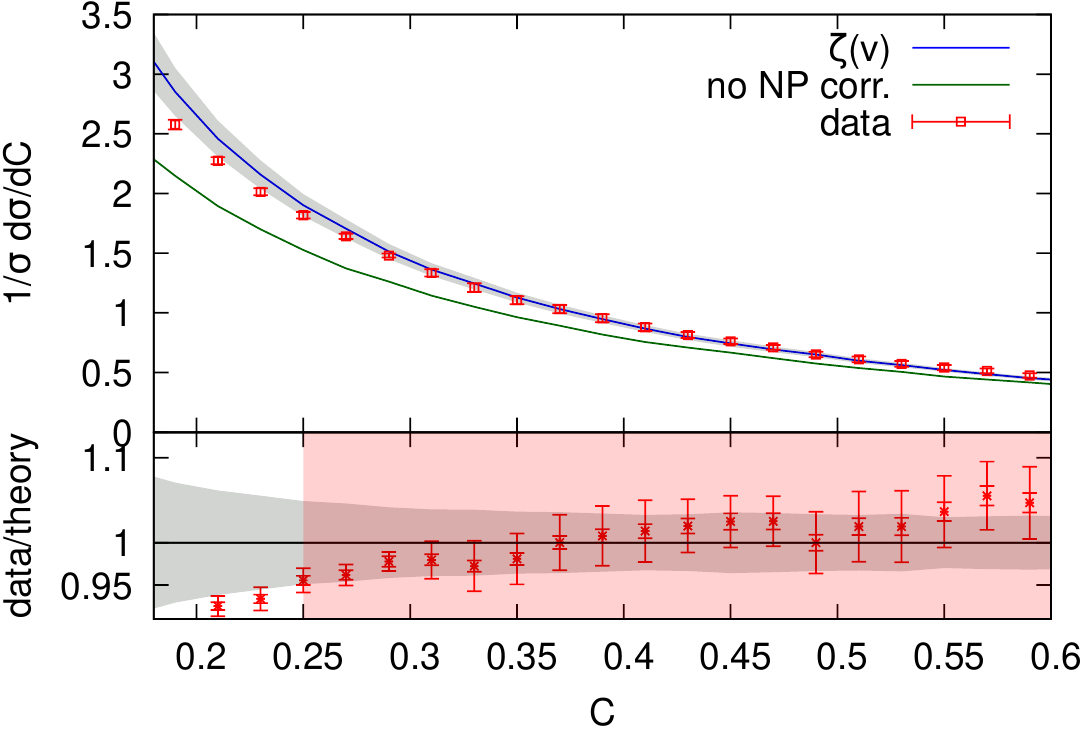}
    \includegraphics[width=0.46\textwidth,page=1]{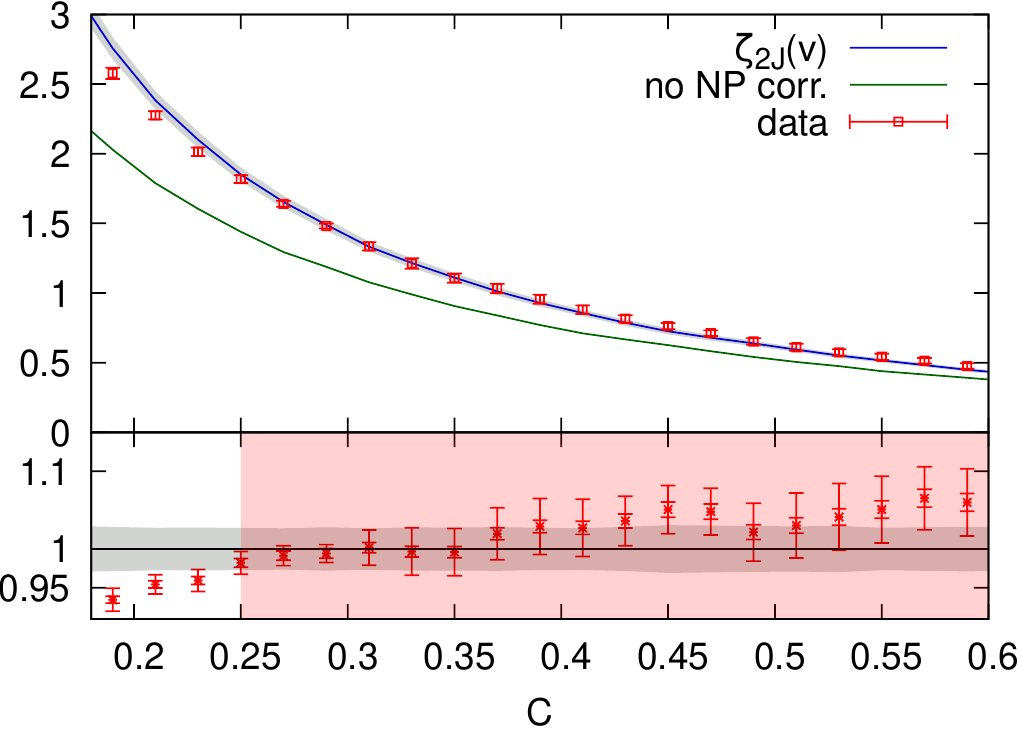}
    \includegraphics[width=0.495\textwidth,page=2]{Figures/fitplot-full}
    \includegraphics[width=0.46\textwidth,page=2]{Figures/fitplot-constzeta-noy}
    \includegraphics[width=0.495\textwidth,page=3]{Figures/fitplot-full}
    \includegraphics[width=0.46\textwidth,page=3]{Figures/fitplot-constzeta-noy}
    \caption{Theoretical predictions compared to data for our default
      setup on the left side, and the default setup with the
      $\zetatj$ functions on the right side.  The gray band represents the
      theoretical errors, while the red bars indicate the experimental
      ones, with the smaller one representing the statistical error,
      and the green lines show the pure perturbative results.
      The highlighted region represents the fit range.
    \label{fig:fitquality}}
  \end{center}
\end{figure}

A more detailed comparison of our default fit with the newly
calculated $\zeta$ functions, and with the $\zetatj$ functions
corresponding to what has been available until now is shown in
Fig.~\ref{fig:as-a0-full-const}.
As mentioned earlier, both fits look plausible, was it not for the fact
that the $\zetatj$ result favours values of $\as$ lower than
the world average. The quality of the fits is displayed in Fig.~\ref{fig:fitquality}.
As one can see, the fit with the full $\zeta(v)$ dependence seems
slightly better, while the one with the $\zetatj$ functions exhibits
some tensions among the different observables. However, on the basis
of the $\chi^2/N_{\rm deg}$ values, both fits are quite acceptable.

\begin{figure}[htb]
  \begin{center}
    \includegraphics[width=0.495\textwidth,page=4]{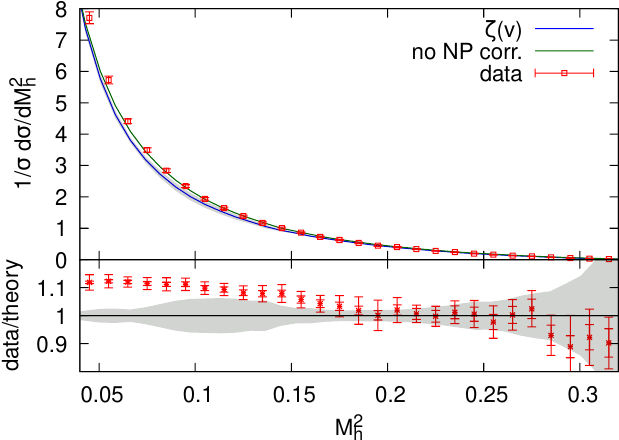}
    \includegraphics[width=0.46\textwidth,page=4]{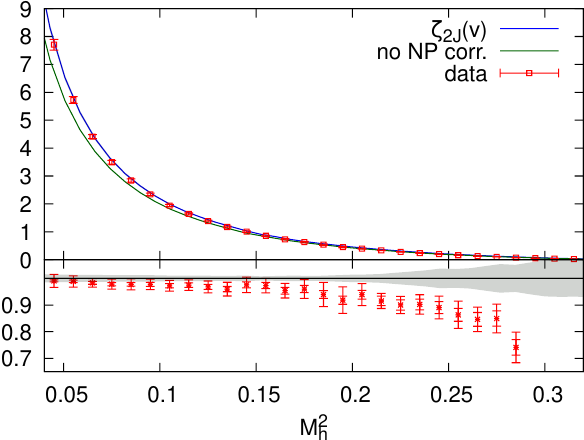}
    \includegraphics[width=0.495\textwidth,page=5]{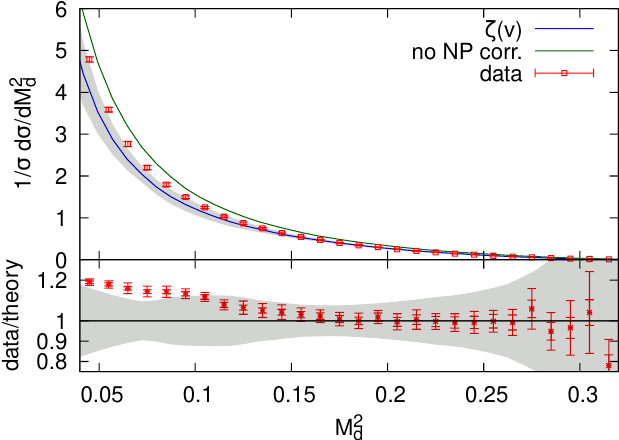}
    \includegraphics[width=0.46\textwidth,page=5]{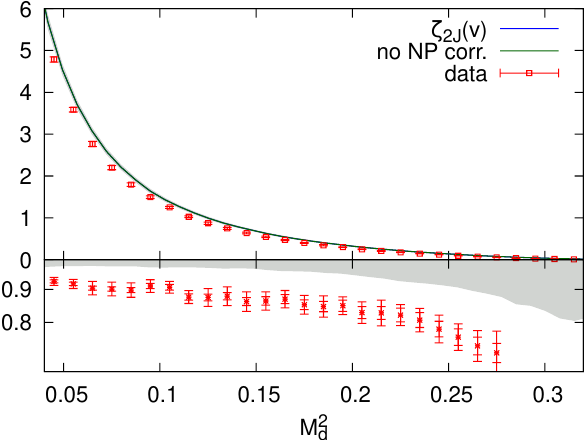}
    \includegraphics[width=0.495\textwidth,page=6]{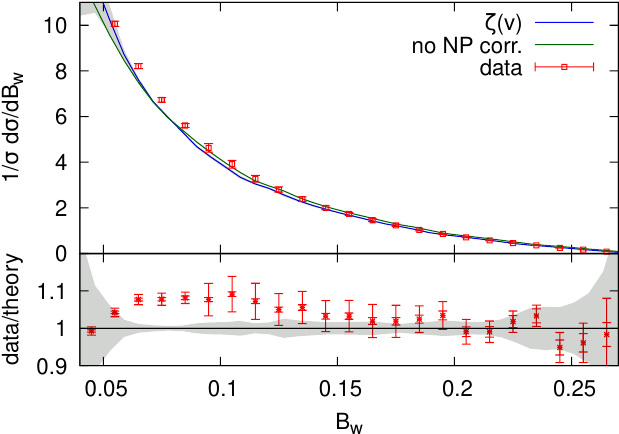}
    \includegraphics[width=0.46\textwidth,page=6]{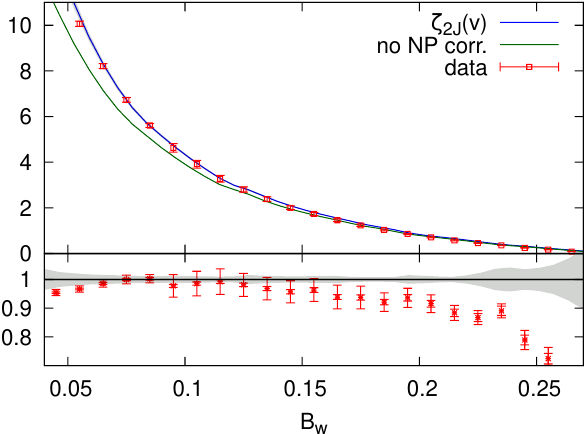}
    \caption{Theoretical predictions compared to data for our default setup
      on the left side, and the default setup with the $\zetatj$ functions on the right side,
      for the  $\Mh^2$, $\Md^2$ and $\Bw$ shape variables. The gray band represents the
      theoretical errors, while the red bars indicate the experimental
      ones, with the smaller one representing the statistical error.
      The green lines show the pure perturbative results. 
    \label{fig:fitother}}
  \end{center}
\end{figure}

It is now interesting to see what happens to the remaining shape
variables, $\Mh^2$, $\Md^2$ and $\Bw$ evaluated with the same
parameters used for our default fits.  The result is displayed in
Fig.~\ref{fig:fitother}.
There we see distinctly that the full $\zeta(v)$ fit works very well
towards the three jet region for all the observables. The $\zetatj$
fit, on the other hand, does not work in the three-jet limit, while
its description of data improves in the two-jet region, with the
noticeable exception of $\Md^2$.

\subsection{On the structure of $\as \lambda/Q$ corrections}
Higher-order corrections to the linear $\lambda$ term are certainly present.
The important issue is whether these corrections are of order $\as(Q)$ or rather
$\as(\lambda)$. In this work we are implicitly assuming that they
are suppressed by a power of $\as(Q)$. We do not have a solid argument to prove
this assumption.
However, by examining the structure of the linear power corrections
near the two-jet limit we gain some insight into how this may actually work.
In fact one can write schematically the correction of order $\lambda$ to a shape
variable $v$ in the two-jet limit as\footnote{This holds for all the observables
  that we are considering with the exception of $\Bw$.}
\begin{equation}
  \left.\frac{\mathd \sigma}{\mathd v}\right|_{\rm \lambda}=N\lambda\left[
    \delta'(v) \zeta_{\rm 2j}+ (\delta'(v)V_1+\delta(v)V_2)\as + \frac{\mathd}{\mathd v}\left(\frac{\mathd \sigma_{q\bar{q}g}}{\mathd v}\zeta(v)\right)
    \right],\label{eq:corrform}
\end{equation}
where the first term is the correction to the leading (two-parton)
configuration, the second term is the virtual correction of order
$\as$, and the third term is the correction we have computed, and
where we implicitly assume some regularization of the $v \to 0$
region.  The derivative of the delta function in the first term is
necessary to guarantee that upon integration in $v$ there are no
linear corrections left at order zero in $\as$, since we know that they are
absent in the total cross section.
The terms $V_1$ and $V_2$ incorporate
corrections where the hard gluon is virtual and the gluer is real or
virtual. In this case, besides the derivative of the
$\delta$-function, we also include an explicit $\delta$-function to
indicate that terms that do not vanish upon integration
in $v$ must exist and are in fact divergent.
We do not include virtual
corrections to the $q\bar{q}g$ process for the exchange of a virtual
gluon of mass $\lambda$, since it was shown in
ref.~\cite{Caola:2021kzt} that these do not lead to linear terms in
$\lambda$.
The absence of linear corrections to the total cross section leads us to
conclude that the integral of the above formula from $v=0$ up to any
finite value of $v$ must be finite.
In fact, if that was not the case, such divergence could not be canceled
when performing the integral in the whole range of the shape
variable. Thus the argument of $\as$
must be taken equal to the hard scale (that in this case is not quite
$Q$, but is related to the typical transverse momentum of the perturbative
gluon that sets the value of $v$). We have thus shown that the singular
contributions of the hard gluon (hard relative to the scale $\lambda$)
in the real emission and virtual exchanges cancel each other also in the
coefficient of the linear term.

The argument given above also suggests a possible way to match
the linear corrections in the three-jet limit to those in the
two-jet limit, that are entangled with resummation effects.
If we recall that the two-jet limit of the functions $\zeta(v)$
for $C$, $\tau$, $\Mh^2$ and $\Md^2$ approach the value $\zeta_{\rm 2j}$,
we could conclude that the part of the last term in the square bracket of
eq.~(\ref{eq:corrform}) that is singular in the two jet
limit must combine with the virtual correction to yield a finite
result. This combined result is precisely what one gets when expanding
in powers of $\as$ the Sudakov form factor, including the shift for
the two-jet non-perturbative correction. Thus, it is tempting to
conclude that the singular part of the last term function should be
combined with the resummation component of the cross section, while
only the regular part should be applied to the 3-jet region. It is
unlikely, however, that this approach will work for observables like
$\Mh^2$ and $\Md^2$, since in their case the limiting value is
approached extremely slowly, and in the first case it has even
opposite sign with respect to the average value of the $\zeta$
function in the fit range. It is however reassuring to see that if we
restrict ourselves to regions far away the two-jet region, all shape
variables are well described with the $\zeta$ functions computed here,
while this is not the case with the values of table~\ref{tab:zeta0}.

\section{Conclusions}\label{sec:conclu}

In this work, we study the effect of power corrections in $e^+e^-$
observables in comparison to data, under the light of the new findings
of refs.~\cite{Caola:2021kzt,Caola:2022vea}, where it was shown that
power corrections can be computed directly in the three-jet
configuration, rather than extrapolating them from the two-jet
region. In refs.~\cite{Caola:2021kzt,Caola:2022vea} these power
corrections were computed for the $C$-parameter and for thrust. Here
we also computed them for the three-jet resolution parameter in the
Durham scheme $y_3$, for the squared mass of the heavy hemisphere
$\Mh^2$, for the squared-mass difference of heavy-light hemispheres
$\Md^2$, and for the wide jet broadening $\Bw$. The observables we
considered are those that can be computed in the approach of
refs.~\cite{Caola:2021kzt,Caola:2022vea}, and that are included in the
ALEPH data of ref.~\cite{ALEPH:2003obs}.

For simplicity we stick to a single data set, and we perform our
calculation using the NNLO results for $e^+e^-$ hadronic observables,
plus the newly computed power corrections. We do not attempt to
include resummation effects. Rather, we stick to ranges of the
observables that are far enough from the two jet region so that no
visible depletion of the resummed result with respect to the
fixed-order one is present.

We stress that in this work we are assuming that the non-perturbative
corrections as estimated according to the results of
ref.~\cite{Caola:2021kzt,Caola:2022vea} are not drastically modified by
the inclusion of soft radiation.  Our argument concerning the two-jet
limit region near eq.~(\ref{eq:corrform}) seems to indicate that this is
not the case. However, we are unable to provide a solid argument for
the three-jet region.

Our main results can be summarized as follows.  First of all, for all
the shape variables that we considered, with the exclusion of the
wide-jet broadening, the function that parameterised the
non-perturbative correction, called $\zeta(v)$, approaches its two
jet-limit value when its argument approaches the two-jet limit value
(set conventionally to $v=0$), as one expects according to simple
physics arguments. However, with the exception of $y_3$, the limit, is
approached only for exponentially small values of the shape variable,
so that, in practice, one sees an effective jump of the function near
$v=0$. This jump is not very important for $C$ and for the thrust
$\tau=1-T$, where it is around 10-20\%{} of the two-jet limit
value. It is instead quite large for $\Mh^2$ and $\Md^2$, where it is
such that the two-jet limit value cannot be considered representative
of the value of the function even very close to the two-jet limit. In
view of these observations, we exclude these observables from our fit,
and also exclude $\Bw$ that is positive and divergent in the two-jet
limit, and is instead negative in the three-jet region.

We thus fitted $C$, $\tau$ and $y_3$, extracting a value for the strong
coupling constant on the $Z$ peak, and for the non-perturbative
parameter $\alpha_0$.  The result of the fits yield a value of $\as$
in acceptable agreement with the world average, although we find that
a number of variations of our procedure can lead easily to differences
of the order of a percent.
Using the same value of $\as$ and
$\alpha_0$, we see that we can describe quite well also the remaining
observables $\Mh^2$, $\Md^2$ and $\Bw$, as long as we remain far
enough from the two-jet limit.  Conversely, with the traditional
implementation of power corrections, good fits to
$C$, $\tau$ and $y_3$ can also be obtained, however
the description of $\Mh^2$, $\Md^2$ and
$\Bw$ in the three-jet region is totally unacceptable.

We stress again that the inclusion of resummation effects in the bulk of the
three jet region leads smaller values of $\as$.\footnote{
  In particular, for fits to the $C$-parameter one finds values of $\as$
  smaller by about ten percent (private communication by P.~Monni).}

We are aware that the present work should only be considered as a
preliminary exploration of the implications of the results of
refs.~\cite{Caola:2021kzt,Caola:2022vea}. In fact, there are few
directions that need further exploration in order to fully exploit
these new results.

First of all, it would be interesting and important
to also include resummation effects in our analysis. Some ideas
regarding this are discussed in the text, suggesting that perhaps the
two-jet limit shift should be applied to the resummed
component of the cross section, while the full $\zeta(v)$ dependent
part should be applied to the finite part. Yet, whether this approach
is sensible also when including resummation effects far from the
two-jet region is a question that needs to be examined more closely,
since for most observables $\zeta(0)$ differs considerably from
$\zeta(v)$ in the three-jet region.

A second direction of improvement regards the choice of the
hadron mass-scheme. Lacking a theoretically sound treatment of this
problem, a possible development would be to see if there is a scheme
that is preferred by data. This in turn would require considering
enough observables that display different behaviour regarding the
mass-scheme choice.

This brings us to consider a third extension of
this work, which is to examine more variables, and find a sufficiently
large set such that the requirements for the applicability of the
results of refs.~\cite{Caola:2021kzt,Caola:2022vea} are met, and such
that their behaviour near the two-jet limit are closer to that of the
thrust and the $C$-parameter. These new variables, could also be
analyzed at present using preserved LEP data~\cite{DPHEP:2015npg}, while waiting
for the beginning of operation of new $e^+e^-$ colliders.

\section*{Acknowledgments}
P. N. would like to thank the Max Planck Institute for hospitality while
part of this work was carried out.
We thank Andrea Banfi, Adam Kardos, Stephan Kluth, Pier Francesco Monni, Silvia
Ferrario Ravasio, Gavin Salam, Hasko Stenzel, Roberto Tenchini, and
Andrii Verbytskyi for useful discussions.

\newpage 
\appendix

\section{Impact of resummation}\label{app:resum}
\begin{figure}[htb]
  \begin{center}
    \includegraphics[width=0.64\linewidth,page=2]{Figures/FO-vs-matched}
    \includegraphics[width=0.64\linewidth,page=1]{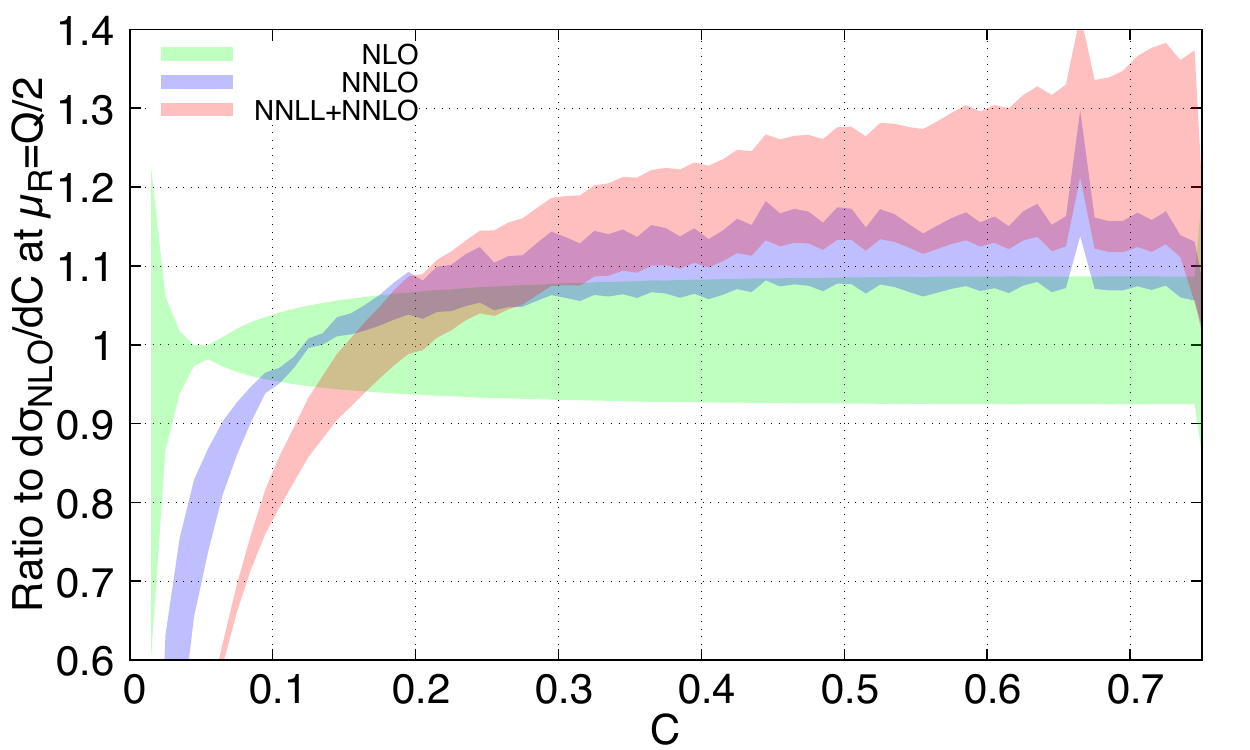}
    \includegraphics[width=0.64\linewidth,page=3]{Figures/FO-vs-matched}
    \caption{Comparison between NLO (green bands), NNLO (blue bands) and NNLO+NNLL
      (red bands) predictions for the thrust (left), $C$-parameter
      (central), $y_{3}$ (right). See
      text for more details.
     \label{fig:FOvsmat}}
  \end{center}
\end{figure}

The fits of $\alpha_s$ carried out in this work rely on fixed order
NNLO predictions, rather than on all-order (NNLL) predictions matched
to fixed order, as computed in
Ref.~\cite{deFlorian:2004mp,Becher:2008cf,Monni:2011gb,Chien:2010kc,Becher:2010tm,Becher:2012qc,Alioli:2012fc,Banfi:2014sua}
for event-shapes and in Ref.~\cite{Banfi:2016zlc} for the Durham
three-jet resolution parameter $y_{3}$.
Although it is customary to include resummation effects also far away
from the two-jet region, in this work we made the assumption that
resummation effects should not be included when the logarithm of the
shape variable is not large. In order to determine a range for the
fit, we thus compare in Fig.~\ref{fig:FOvsmat} NNLO and
NNLO+NNLL predictions for the thrust variable $\tau=1-T$, the
$C$-parameter, and the Durham three jet resolution variable $y_{3}$
and exclude in our fits the regions where matched predictions clearly
depart from the fixed order.
Each plot shows the ratio to the NLO prediction obtained with central
renormalization scale $\mu_{R,0}=Q/2$. The green band shows the
uncertainty of the NLO and the blue band of the NNLO, and are
obtained by varying $\mu_R$ up and down by a factor two around the
central value. 
For the NNLO+NNLL matched predictions we fix our default
setup as follows: we set the central renormalization scale to
$\mu_{R,0}=Q/2$, the resummation scale to $\mu_{Q,0}=Q/2$, we use the
modified logarithm $L=1/p\ln\left(1/v^p-1/v_{\rm lim}^p+1\right)$,
where $v_{\rm lim}$ denotes the kinematic limit of the event shapes,
with p=3, and we use the log-R matching scheme (see
e.g. ref.~\cite{Banfi:2014sua}).
The uncertainty band is then obtained as follows.  Around the above
described default setup, we vary, one at the time,
$\mu_{R,0/2}\le\mu_{R}\le 2\mu_{R,0}$, $\mu_{Q,0}/2\le\mu_{Q}\le
2\mu_{Q,0}$, we vary $p$ to $p=2$ and $p=5$, and, finally, we use the
R-matching scheme. This gives a total of eight matched
predictions. The red uncertainty band shown in Fig.~\ref{fig:FOvsmat}
is obtained by taking the envelope of all these predictions.

The onset of resummation effects is signalled both by a drop of the
distribution of the resummed result and by an increase of the NLO
result with respect to the NNLO one. We choose the lower bound of our
fit ranges to be to the right of this region. 
Furthermore, for the three observables used in the fit we observe the
following features: for the thrust, the uncertainty bands of
the NNLO and matched predictions overlap, with the resummation band
being a few percent higher, which would lead to slightly smaller
values of $\alpha_s$. For the $C$-parameter one observes a somewhat
similar behaviour. However, the difference between the center of the
resummed and NNLO bands now reach up to 10\% and the resummed
band has a slightly different shape compared to the NNLO
one. For $y_{3}$ one observes small effects, at the level of a 2\%,
however in this case the uncertainty bands do not overlap since the
NNLO band is extremely small. 
From all three plots it is also clear that the difference between
NNLO and matched predictions does not vanish even for large
values of the observables. This is due to the fact that, even with the
modified logarithms, the resummation is not switched off fast enough
even close to the end-point of the distributions.

From the figures it can be seen that in the case of the
thrust, the resummed prediction seems to follow the trend of the NLO
and NNLO corrections, possibly approximating higher-order results if
they follow the same trend. However, in the case of the $C$-parameter
the resummed result has a slope that is not present in the NLO and
NNLO results. Furthermore, in the case of $y_3$, the trend is to have
the NNLO distribution smaller than the NLO one, while the resummed
result is larger.
In conclusion, although it has become common practice, we see no
reason in principle to include resummation effects also in the
three-jet region.

\bibliographystyle{JHEP}
\bibliography{Shapes.bib}
 
\end{document}